\documentclass[12pt]{iopart}
\usepackage{latexsym}
\usepackage[T1]{fontenc}
\usepackage[utf8x]{inputenc}
\usepackage{latexsym}
\usepackage{times}
\usepackage{graphicx,amsfonts,amsbsy,amssymb,amsthm}
\usepackage{amsopn}
\usepackage{color}
\usepackage{cite}
\usepackage{hyperref}

\def\={\;=\;} \def\+{\,+\,}

\begin {document}
\title{The Kronig-Penney model in a quadratic channel with  $\delta $ interactions.  II : Scattering approach. }
\author{Uzy Smilansky}
\address{Department of Physics of Complex Systems, Weizmann Institute of Science, Rehovot 7610001, Israel}
\address{Department of Mathematical Sciences, University of Bath, Bath BA27AY UK.}
\date{\today}
\begin{abstract}
The main purpose of the present paper is to introduce a scattering approach to the study of the  Kronig-Penney model in a quadratic channel with  $\delta$ interactions, which was discussed in full generality in the first paper of the present series. In particular, a secular equation whose zeros determine the spectrum will be written in terms of the scattering matrix from a single $\delta$. The advantages of this approach will be demonstrated in addressing the  domain with total energy $E\in [0,\frac{1}{2})$, namely, the energy interval where, for under critical interaction strength, a discrete spectrum is known to exist for the single $\delta$ case. Extending this to the study of the periodic case reveals quite surprising behavior of the Floquet spectra and the corresponding spectral bands. The computation of these bands can be carried out numerically, and the main features  can be qualitatively explained in terms of a semi-classical framework which is developed for the purpose.
\end{abstract}
\section{Introduction}
This manuscript is the second in a series of two papers dedicated to the same subject - the study of the Kronig-Penney model in a quadratic channel with periodic $\delta$ potentials coupled linearly to the transversal degree of freedom. The first paper \cite {Italo I} (to be referred as (I) in the sequel), provides the background, the motivation and the scope of the series, together with a brief  review of the relevant literature.

However, in order to introduce the notation and to render this paper self contained, the system to be studied is briefly described in the following paragraph.

The Schr\"odinger operators to be addressed here, expressed in the $(x,q)$ representation  ($(x,q)\  \in\ {\mathbb R}^2$) are defined by
 \begin{eqnarray}
  \label {schroedinger}
 H& = & H_0 +V_L(x,q) \nonumber \\
 H_0 &=& -\frac{\partial^2\ }{\partial x^2}+h  =  -\frac{\partial^2\ }{\partial x^2} +\frac{1}{2}( -\frac{\partial^2\ }{\partial q^2} + q^2 ) \nonumber  \\
 V_L(x,q)&=&\lambda q\ \sum_{m\in {\mathbb Z}} \delta(x-L m).
 \end{eqnarray}
 The spectrum of $H_0 $ denoted by  $\sigma(H_0)$  is absolutely continuous, with multiple degeneracies  and  supported on the positive real axis with $E\ge \frac{1}{2}$. Denote the eigen-functions of the harmonic oscillator operator $h$ by $f_n(q)$ with eigenvalues  $(n+\frac{1}{2}) $  for $ n \in \mathbb {N}^0 $. Then,    $\sigma_0$ is of  multiplicity
$n(E) = \lfloor E-1/2\rfloor$ so that $E=(n+\frac{1}{2} + k_n^2)$, corresponding to the eigen-functions $f_n(q)e^{\pm i k_n x } $ with $0 < n < n(E)$ and   with $k_n\in  {\mathbb R}^+ $. The branch of $\sigma(H_0)$ for a particular value of  $n$ is referred to as a "mode".  The  eigen-functions  are extended on the $x$ axis, but restricted   by the quadratic (harmonic) potential to remain in a channel close to the $x$ axis.
 $H $ is obtained by adding to  $H_0 $ a periodic potentials $V_L(x,q)$  where $L$ is the distance between successive locations of the $\delta (x)$  potentials.

This system belongs to a family of models which address various problems that arise when a single or several  delta interactions are added to $H_0$.
(see representative references- \cite{irrev,SolomyakUS,s1,s5,s2,s3,ex1,ex2,ex3,ex4,Italo1,Italo2})  Here we focus on the coupling operator $\lambda q \delta(x)$ and introduce a scattering approach which is similar to the one  commonly used in the study of the Schr\"odinger operator on metric graphs  (alias "quantum-graphs").  There, the building-block of the theory is the scattering matrix at each single vertex,  the dynamics  is strictly one dimensional and the wave-number is not changed in the scattering. Here, a "vertex" stands for the  localized coupling operator, where the energy is shared and interchanged between the motion along the $x$ (longitudinal)  and the $q$  (transversal) directions.  This  enriches the variety of phenomena which can be discussed in terms of graph models,  and at the same time poses challenging difficulties for the mathematical treatment. Another article in the present volume addresses multi-mode graphs from a different point of view \cite{SvenUS}.

The paper is arranged in the following way.
In the  rest of the introduction, the scattering approach will be applied to the original Kronig-Penney model with $\delta$ potential  \cite{kporiginal}.  This  one-dimensional problem is a linear quantum graph with vertices of degree 2. This is why it serves as a good starting point for  the study of  the two dimensional problem using the scattering approach.

The  scattering matrix from a single $\delta$ interaction will be introduced and studied  in section (\ref {first}). It will be used for writing a spectral secular equation which will be studied in detail for the energy domain $E\in (0,\frac{1}{2})$ where a point spectrum is known to exist.   The derived scattering matrix will be used in section (\ref {second}) to  study of the Floquet spectra and the spectral bands of the  generalized Kronig-Penney model (\ref {schroedinger}), with particular attention to the same energy interval, namely $E\in (0,\frac{1}{2})$.

\subsection{ A primer- Scattering approach to the  Kronig-Penney model }
This model is discussed in numerous publications and books, see e.g., \cite{{kpbook}} and reference cited therein. It is brought here to introduce the scattering approach by its application to a simple and familiar problem.
\subsubsection{The scattering matrix from a single delta potential}
The Schr\"odinger operator with the $\delta$  potential centered at $x=0$ reads
\begin{equation}
H(x)= -\frac{\partial^2\ }{\partial x^2}   + \lambda \delta (x) \ \ {\rm for }\ \  x \ \ \in\mathbb{ R}
\label{schroekp}	
\end{equation}
Write
\begin{eqnarray}
\Psi(x)&=& \Psi^{(a)}(x) \frac{1-{\rm Sign}[x]}{2}+ \Psi^{(b)}(x) \frac{1+{\rm Sign}[x]}{2} \\
\Psi^{(a)}(x )&=&   \frac{1}{\sqrt{| k |}}(a ^+ e^{+ik x} +a ^- e^{-ik x} )    \\
\Psi^{(b)}(x )&=&   \frac{1}{\sqrt{| k |}}(b ^+ e^{+ik x} +b ^- e^{-ik x} )  \nonumber
\label{wfunctkp}
\end{eqnarray}
Where the spectrum of $H(x)$ is denoted by $E=k^2$ and for $E<0$ $k=+i\kappa$.  The boundary conditions at $x=0$ are
continuity of $\Psi(x)$ , and a discontinuity
\begin{equation}
\lim_{\epsilon\rightarrow 0^+}\left (\frac{ \partial \Psi^{(b)}(x=+\epsilon)}  {\ \ \ \partial x\ \ }-
\frac{ \partial \Psi^{(a)}(x=-\epsilon)}  {\ \ \ \partial x\ \ }      \right ) = \lambda \Psi(x=0)\ .
\label{bcpk}
\end{equation}
These conditions provide the relations
\begin{equation}
a^+ + a^- =b^+ + b^-\ , \ \   {\rm and }\ \  (b^+ - b^-) -(a^+ - a^-) =\frac{\lambda}{ik}(a^+ + a^-)  .
\label{bckp1}
\end{equation}
The amplitudes $ a^+$  (and  $b^-$) are the amplitudes of waves which impinge on the potential from resp. the left (right) directions, while the corresponding scattered amplitudes, are $ a^-$  (and  $b^+$). The incoming and outgoing pairs are related by a scattering matrix which can be easily computed using (\ref {bckp1}) above:
\begin{eqnarray}
\left (  \begin {array}{l}
a^-\\
b^+
\end{array}
\right )\ = S(k)
 \left (  \begin {array}{l}
a^+\\
b^-
\end{array}
\right )\ \nonumber \\
S(k)=\frac{1}{1+i\frac{\lambda}{2k}}
\left (  \begin {array}{ll}
\frac{\lambda}{2ik} & 1\\
1 & \frac{\lambda}{2ik}
\end{array}
\right ) \ =-\left (  \begin {array}{ll}
1 & 0\\
0 & 1
\end{array}
\right ) +\frac{1}{1+i\frac{\lambda}{2k}}
\left (  \begin {array}{ll}
1 & 1\\
1 & 1
\end{array}
\right )
\label{scatmatkp}
\end{eqnarray}
The unitarity of $S(k)$ for real $k$ guarantees conservation of flux, namely, $|a^-|^2+|b^+|^2=|a^+|^2+|b^-|^2$.
Also, The spectrum of $S(k)$ consists of $\{ -1, \frac{1-i\frac{\lambda}{2k}}{1+i\frac{\lambda}{2k}}\}$. Hence  $\det S(k) = -\frac{1-i\frac{\lambda}{2k}}{1+i\frac{\lambda}{2k}}$ has a pole at $k=i\kappa$ so that when $\lambda<0$ the pole is at $\kappa = |\lambda|/2$ which corresponds to a wave function which decays to $0$ as $|x| \rightarrow \infty$. Thus, the $S(k)$ matrix stores the entire information about the wave functions and the corresponding  spectrum of the operator (\ref {schroekp}).


\subsubsection{The periodic delta potential - the Kronig-Penney model }
\label {kpmodel}
The Schr\"odinger operator with periodic $\delta(x)$  potential reads
\begin{equation}
H_{KP}= -\frac{\partial^2\ }{\partial x^2}   + \lambda \sum_{n\in \mathbb {Z}} \delta (x-nL) \ \ {\rm for }\ \  x \ \ \in\mathbb{ R} \ .
\label{schroperkp}	
\end{equation}
The solution is obtained by considering the spectrum of (\ref {schroperkp}) restricted to the unit-cell
$-\frac{L}{2} +\epsilon \le x \le \frac{L}{2}-\epsilon$ with the boundary conditions (\ref {bcpk}) at $x=0$, augmented by the requirement that in the limit $\epsilon \rightarrow 0^+$, and for all $|\omega |\le \pi$
\begin{equation}
\hspace{-20mm}
 \Psi^{(b)}(\frac{L}{2}-\epsilon)=
e^{i\omega}\Psi^{(a)}(-\frac{L}{2}+\epsilon)
\ \ {\rm and  } \ \  \frac{ \partial \Psi^{(b)}( \frac{L}{2}-\epsilon)}  {\ \ \ \partial x\ \ }=e^{i\omega}
\frac{ \partial \Psi^{(a)}(-\frac{L}{2}+\epsilon)}  {\ \ \ \partial x\ \ }\ .
\label{bckp}
\end{equation}
These conditions can be summarized by the requirement that
\begin{eqnarray}
\left (  \begin {array}{l}
b^+\\
b^-
\end{array}\right )=
\left (  \begin {array}{ll}
e^{i(\omega -kL)} & 0\\
0 & e^{i(\omega +kL)}
\end{array}
\right )
\left (  \begin {array}{l}
a^+\\
a^-
\end{array}
\right ) \ .
 \label{blochkp}
\end{eqnarray}
The scattering on the potential  $\lambda \delta(x) $ impose    (\ref {scatmatkp}) on the amplitudes $a^{\pm}, b^{\pm}$. Together, they require
 \begin{eqnarray}
 \left (  \begin {array}{l}
 a^+\\
 a^-
 \end{array}\right )=e^{ikL}
 \left (  \begin {array}{ll}
0& e^{i \omega } \\
 e^{-i \omega }&0
 \end{array}
 \right ) S(k)
 \left (  \begin {array}{l}
 a^+\\
 a^-
 \end{array}
 \right ) \ .
 \label{conskp}
 \end{eqnarray}
This is a set of two homogeneous linear equations in two unknowns, which are consistent only at the zeros of the {\it secular function} $\zeta_{kp}(k;\omega)$, with
\begin {equation}
\zeta_{kp}(k;\omega) =\det\left  [I- e^{ikL}
\left (  \begin {array}{ll}
0& e^{i \omega } \\
e^{-i \omega }&0
\end{array}
\right ) S(k)\right  ] =0 \ .
\label {zetakp}
\end{equation}
For each $\omega\ $ this equation is satisfied by a discrete set of spectral points $\left \{k_n(\omega)\right  \}_{n=1}^{\infty}$. $k_n(\omega)$ form the network of Floquet spectral functions.  The spectral bands are defined by the projection of each of the $k_n(\omega)$ on the $k$ axis.  The bands support the continuous spectrum of $H$.
\begin{figure}
 	\includegraphics[width=.50\textwidth]{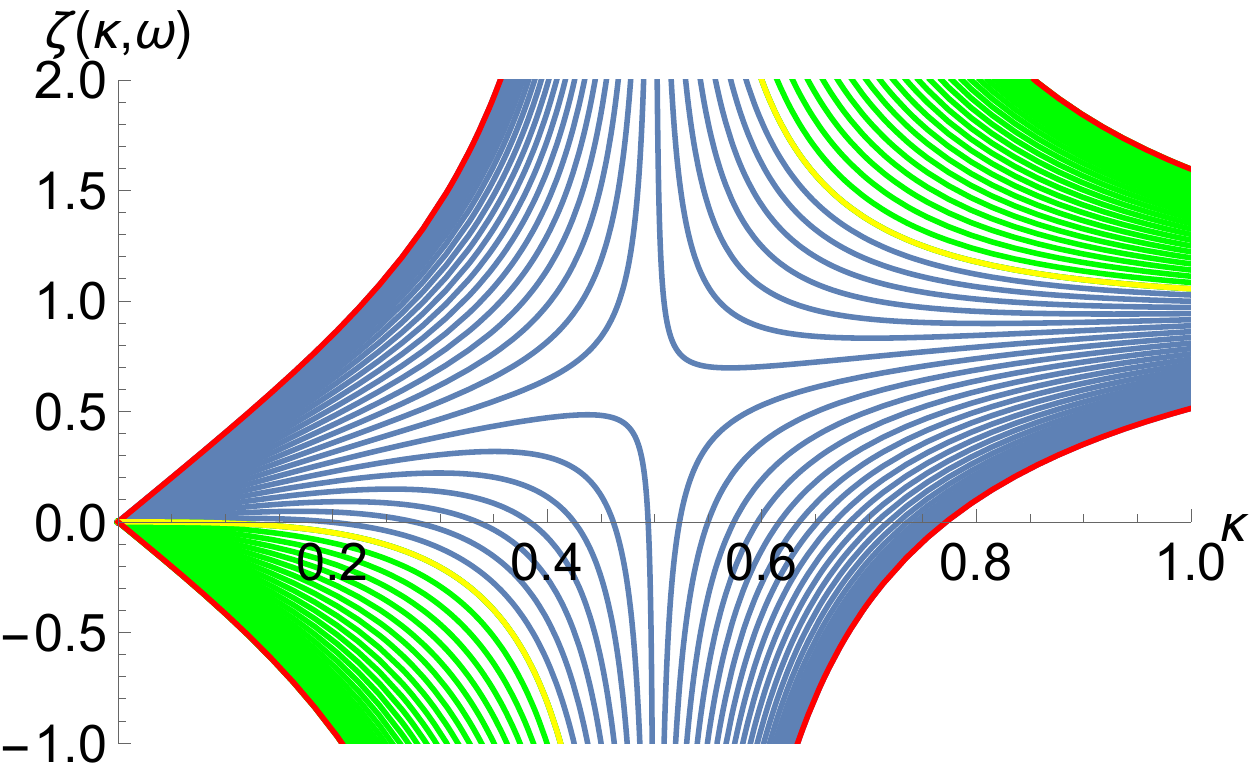}\hfill
 	\includegraphics[width=.50\textwidth]{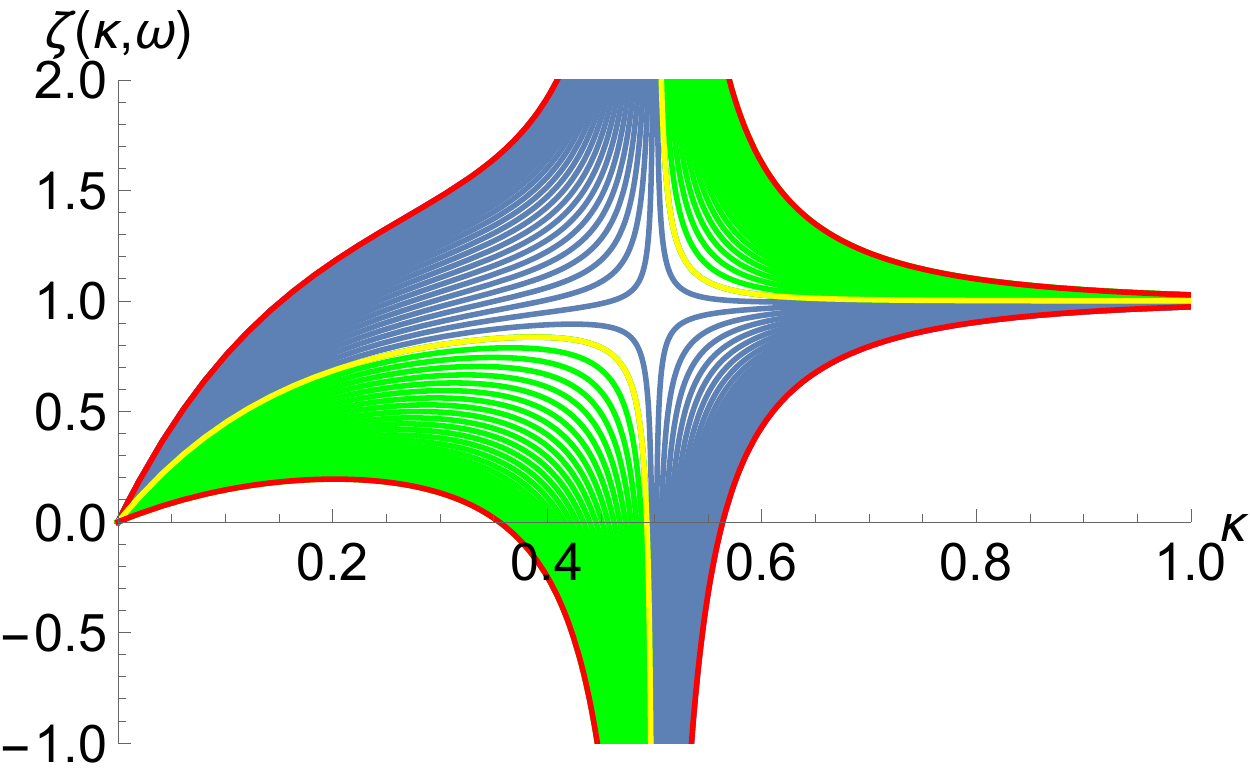}\hfill
 	\includegraphics[width=.50\textwidth]{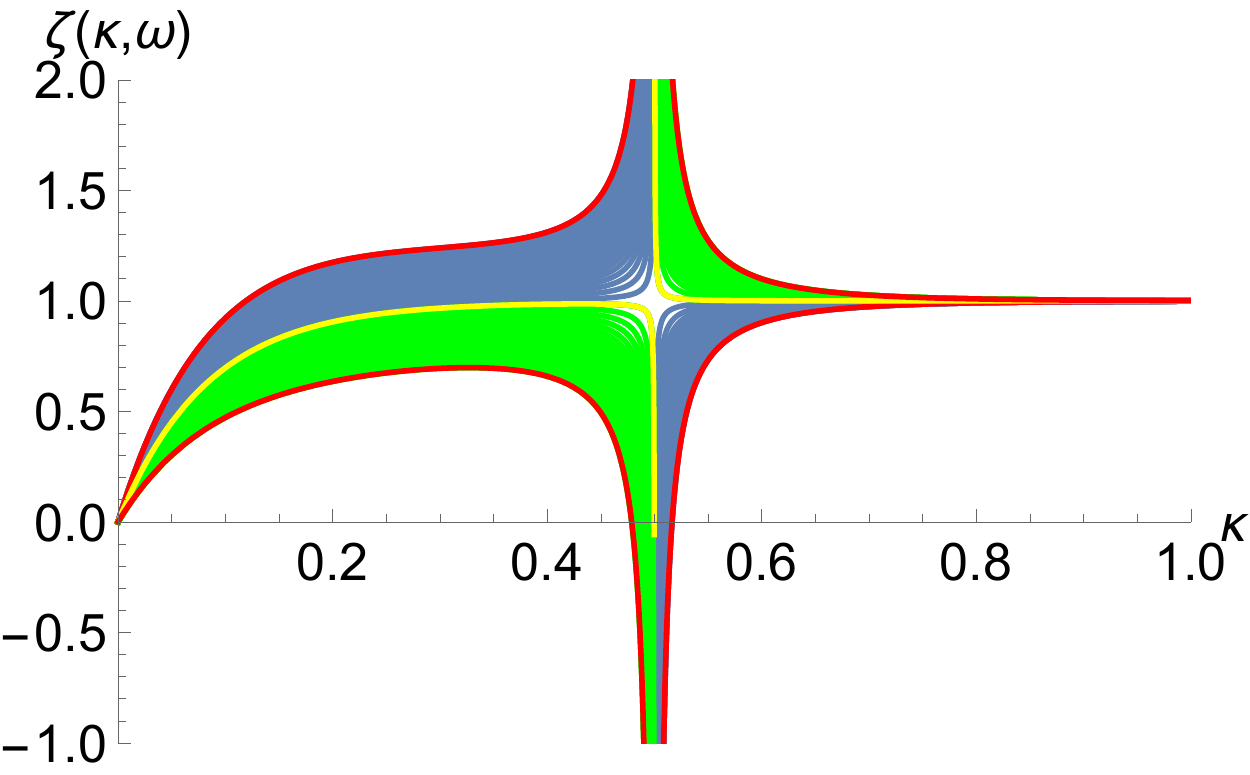}\hfill
 	\includegraphics[width=.50\textwidth]{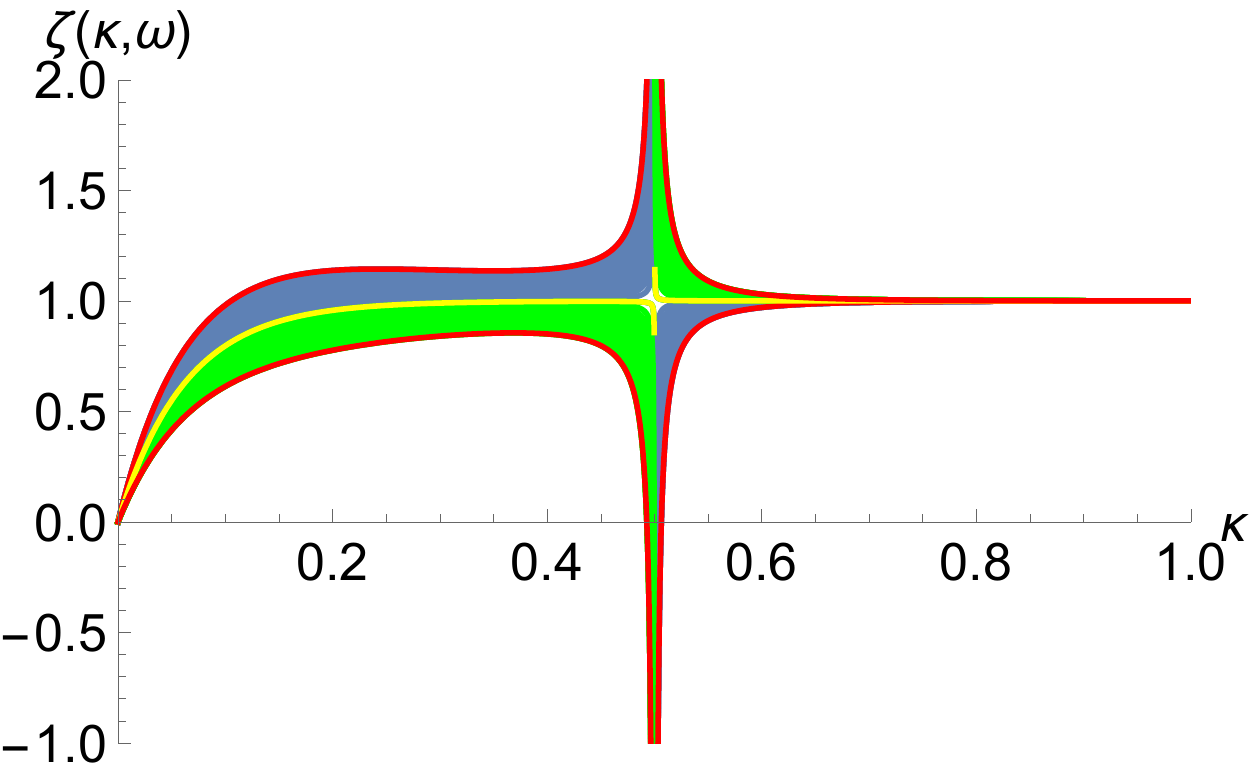}
 	\caption{$\zeta (\kappa;\omega)$, as a function of $\kappa$ for  $|\lambda|=1$ and $L=2,5,8,10$, from top left to bottom right. Each line represents a different value of $\omega$, the blue and green colors are for negative or positive values of $\cos \omega$\ . }
 	\label{fig0}
 \end{figure}
 In the following the discussion will focus on the band structure in the negative  $E$ domain which exists only when $\lambda <0$.  This is done in order to provide a background for the  study of the Floquet spectrum for the periodic operator (\ref {schroedinger}) in the range $0 <E<\frac{1}{2}$.

Substituting $k=i\kappa$ in (\ref{zetakp} ) one gets
\begin{eqnarray}
\label {zetakappa}
\zeta(\kappa;\omega)&=&\frac{2 e^{-\kappa L}}{ 1-\frac{|\lambda|}{2\kappa} }(  \cosh\kappa L-\frac{|\lambda|}{2\kappa  }\sinh \kappa L- \cos\omega )
\nonumber \\
 &=&-\kappa\frac{4}{|\lambda|}(1-\frac{|\lambda| L}{2}- \cos\omega) +\mathcal{O}(\kappa^2)\ .
\end{eqnarray}
Figure (\ref{fig0}) shows $\zeta(\kappa; \omega)$, where each continuous blue (green resp.)  line corresponds to one out of 21 values of  $\omega$ in the range $ 0\le \omega \le \frac{\pi}{2}$, ($ \frac{\pi}{2}\le \omega \le  \pi $ resp.) . The frames  correspond to different  values of the parameters $L$ and $ |\lambda|=1$.

The spectral bands $-\kappa^2 (\omega)$ in the negative energy domains, are the intervals of $\kappa$  where the $\zeta(\kappa,\omega)$ lines intersect the $\kappa$ axis.    There are two Floquet spectra.  One is concentrated  at the  threshold $\kappa=0$. As is evident from  the second line of (\ref{zetakappa}) it is supported by the entire $\omega$ domain $(0,\pi)$. The other is centered at $\omega= \arccos[e^{-\frac{|\lambda|L}{2}}]$ which approaches $\frac{\pi}{2}$ exponentially with $L$. The Floquet spectra are narrowed down in the limit of large $L$, as one expects since the eigenfunctions at each unit cell decay exponentially. Note: the Floquet spectra are not always  supported on the entire allowed $\omega$ interval: the top most green band of $\zeta$ functions never intersect the $\kappa$ axis. The full blue band of $\zeta$ functions intersects the line only for low values of $L$ (see upper left frame corresponding to $L=2$. The dependence of the Floquet spectra on the parameter $L$ becomes much more complicated for the analogous case for the periodic operator (\ref {schroedinger}). This will be discussed at length in section (\ref{sectionkp}).

\vspace{2mm}

\section{A Single Vertex Coupled to the Quadratic Potential }
\label{first}

 Following the practice in quantum graphs, and the analogue example computed for the one-dimensional Kronig-Penney model  (section ( \ref {kpmodel} ) ), the scattering matrix  associated with a single vertex is computed. The scattering approach will be illustrated by discussing the pure point spectrum in the energy domain $\frac{1}{2} >E>0$ for $\Lambda <1$,  retrieving the known results of \cite {s1,s2,s3,s5,ex1,ex2} from a different perspective.

 \subsection {The scattering matrix for a single $\delta (x)$  potential}

The Schr\"odinger operator to be considered in the present section is
 \begin{eqnarray}
 \label {schroedsingr}
 H_1  =   -\frac{\partial^2}{\partial x^2 }   +\frac{1}{2}( - \frac{\partial^2}{\partial q^2} + q^2 )
+\lambda q\  \delta(x)   \ \ , \ \  (x,q)\in {\mathbb R}^2 .
 \end{eqnarray}
It is augmented by boundary conditions at $x=0$  which require that $\Psi (x,q)$  is continuous at $x=0$,  but its derivative by $x$ at $x=0$ is discontinuous: :
\begin{equation}
\lim_{\epsilon\rightarrow 0^+}\left (\frac{\partial\ }{\partial x }\Psi (x=+\epsilon,q) -
 \frac{\partial\ }{\partial x } \Psi (x=-\epsilon,q)   \right ) = \lambda q \Psi(x=0,q)\ , \ \ \ \forall q \ .
\label{bcsingle}
\end{equation}
The wave function which satisfies
\begin{equation}
H_0\Psi (x,q) = \left [-\frac{\partial^2}{\partial x^2 } \   +\frac{1}{2}( - \frac{\partial^2}{\partial q^2 }  + q^2 ) \right ] \Psi (x,q)=  k^2 \Psi (x,q)
\label{hzero}
\end{equation}
away from $x=0$ is expanded by writing
\begin{eqnarray}
\Psi(x,q)&=& \Psi^{(a)}(x,q) \frac{1-{\rm Sign}[x]}{2}+ \Psi^{(b)}(x,q) \frac{1+{\rm Sign}[x]}{2} \nonumber \\
\Psi^{(a)}(x,q))&=& \sum_{m=1}^{\infty}  \frac{1}{\sqrt{| k_m|}}(a_m^+ e^{+ik_mx} +a_n^- e^{-ik_mx} ) f_m(q) \nonumber \\
\Psi^{(b)}(x,q))&=& \sum_{m=1}^{\infty}  \frac{1}{\sqrt{| k_m|}}(b_m^+ e^{+ik_mx} +b_m^- e^{-ik_mx} )f_m(q) ,
\label{wfunction}
\end{eqnarray}
where,  $f_m(q)$ stands for the eigenfunction of $h=\frac{1}{2}( - \partial_q^2 + q^2 ) $ with eigenvalue $(m+\frac{1}{2})  \  , \  m\in \mathbb{N}^{0} $. $k_m^2 \in \mathbb{R}$ is the eigenvalue of the $x$ Laplacian so $k^2 = k_m^2 +(m+\frac{1}{2}) $ is the eigenvalue of $ H_0$ (\ref{hzero}). As long as $k^2\ge
(m+\frac{1}{2}), \ k_m=  + \sqrt { k^2-(m+ \frac{1}{2})}$ and $k_m$ is the wave number in the $n$'th conducting mode.
If $k^2< (m+\frac{1}{2}), \ k_m=  + i\sqrt {(m+ \frac{1}{2})- k^2}  =i\kappa_m$ and $\kappa_m$   is the rate of the exponential decay or growth of the wave functions in the corresponding evanescent mode.

The difference between the presentations here and in  paper (I) of  this series is in the normalization factors $\frac{1}{\sqrt{k_n}}$ which appear here and are missing in (I). They are included here because they normalize the wave functions in the conducting modes to have a unit flux at infinity. This is necessary in order to render the scattering matrix unitary.

To facilitate the notations,  bold-face letters will be used to denote infinite vectors so that e.g., ${\bf a^{\pm}}$ stands for the column vector $(a^{\pm}_0,a^{\pm}_1,\cdots)^ {\top}$, and bold-face capital letters stand  for the  corresponding matrices, so that e.g., the infinite identity matrix is denoted by ${\bf I}$.

Applying the continuity and jump conditions at $x=0$ to the expansion (\ref {wfunction}), multiplying by $f_n(q)$ and integrating over $q$ results in
\begin{equation}
  a_n^+ +  a_n^-  =  b_n^+ +  b_n^- \ .
\label{contab}
\end{equation}
Making use of $\int_{\-\infty}^{\infty} f_n(q) q f_{m}(q) {\rm d}q=\delta_{m,n\pm1} {\sqrt \frac{n+\frac{1}{2}\pm \frac{1}{2}}{2}} $   and denoting $\Lambda=\frac{\lambda}{\sqrt{2}}$ results in ,
\begin{eqnarray}
\label{jacob1}
&i&\frac{k_n}{\sqrt { |k_n|}}[(b_n^+ - b_n^-)-(a_n^+ - a_n^-)   ] = \nonumber \\
& &\Lambda\left [{\sqrt\frac{n+1}{|k_{n+1} |}} (b_{n+1}^++b_{n+1}^-)+
{\sqrt\frac{n\ }{|k_{n-1} |}} (b_{n-1}^++b_{n-1}^-) \right] \ .
\end{eqnarray}
Defining the  tridiagonal matrix  $ {\bf J_0}(k) $ by
\begin{equation}
\left ({\bf J_0 }(k)\right )_{n,n'} = \
i\Lambda\frac{|k_n|}{k_n} \left [
{\sqrt\frac{n+1}{|k_n k_{n+1} |}} \delta_{n',n+1}
+
{\sqrt\frac{n\ }{|k_n k_{n-1} |}} \delta_{n',n-1}^+  \right]
\label{j0}
\end{equation}
The equations (\ref {contab},\ref {j0} ) can be conveniently written as
\begin{eqnarray}
{\bf a}^+ +  { \bf a}^-  &=&  { \bf b}^+ + {\bf  b}^- \nonumber \\
{\bf a}^+ -  { \bf a}^- &=& ({\bf J_0}(k)+{\bf I}){ \bf b}^+ +({\bf J_0}(k)-{\bf I}){ \bf b}^- \ .
\label{pres}
\end{eqnarray}
These equations are rearranged by introducing formally a scattering matrix ${\bf S}(k)$
\begin{eqnarray}
\left ( \begin {array} {l}
{\bf a}^{-}\\
{\bf b}^{+}\\
\end{array}\right )=
{\bf S}(k)
\left ( \begin {array} {l}
{\bf a}^{+}\\
{\bf b}^{-}\\
\end{array}\right )\  ,
\label{rearanger}
\end{eqnarray}
where the scattering matrix is
\begin{eqnarray}
{\bf S}(k)=- \left (
\begin {array} {cc}
{\bf I} &0 \\
0 &{\bf I}\\ \end{array}\right )
+\left [{\bf I} +\frac{1}{2}{\bf J_{0}}(k) \right ]^{-1}
\left (  \begin {array} {cc}
{\bf I} &{\bf I} \\
{\bf I} &{\bf I}\\ \end{array}\right )\ .
\label{smatrixq}
\end{eqnarray}
Note that the scattering matrix as defined above acts on vector pairs  which are  written as two component vectors as in ( \ref {rearanger}). Note also the structural resemblance between (\ref {smatrixq}) and (\ref {scatmatkp}). The scattering matrix as defined above is an infinite matrix, whose proper definition is deferred to later on in this section.

The single delta model was discussed in the past by several authors \cite{s1,s2,s3,s5,ex1,ex2}, and its spectral  as well as its dynamical \cite{Italo1,Italo2}  and scattering properties \cite{irrev}  were elucidated. Here, a different approach to the study of its spectrum will be displayed, with special attention to the discrete spectrum in the low energy domain $0 <k^2 < \frac{1}{2}$ where for $\Lambda <1$ the point spectrum exists \cite{s2,s5,ex4}, and depends very delicately of the interaction strength $\Lambda$.

\subsection{A secular equation for the point spectrum in the low energy domain}
In the spectral domain $0 <k^2 < \frac{1}{2}$ the momenta $k_n$ for all $n\in {\mathbb N}^0$ are pure imaginary and therefore the amplitudes which correspond to the exponentially increasing functions when $|x|\rightarrow \infty $ must vanish identically. Inserting  ${\bf a^+}={\bf b^-}=0$ in (\ref {pres}) one obtains a condition for the existence of a non trivial solution of
\begin{equation}
(  2{\bf I } +{\bf J_0}){\bf b}^+ =0
\label{cond1}
\end{equation}
namely,
\begin{equation}
  \zeta(k) = \det(2{\bf I} +   {\bf J_0})  =0.
\label{seceq}
\end{equation}
 which defines the secular function $\zeta(k)$  whose zeros are the spectral points at which there exist a  non trivial solution for (\ref {cond1}).
Note that the above condition implies that the discrete spectrum coincides with poles of the $S$ matrix elements
in the domain $0 <k^2 < \frac{1}{2}$.

In order to compute the secular equation, we consider the determinants of the principal minors of dimension $n+1$
\begin {eqnarray}
\label{det}
\hspace{-20mm}
& & D_n(E,\Lambda)=\\
\hspace{-20mm}
& &
\det \left (
\begin {array} {cccccc}
2& \Lambda \sqrt {\frac{1}{\kappa_0\kappa_1}} & 0 &0&\cdot  &0  \\
 \Lambda \sqrt { \frac{1}{ \kappa_1\kappa_0 }}&2& \Lambda \sqrt{ \frac{2}{\kappa_1\kappa_2 }}&0&\cdot &0 \\
0& \Lambda \sqrt { \frac{2}{\kappa_2\kappa_1 }}&2& \Lambda \sqrt{ \frac{3}{\kappa_2\kappa_3 }}&\cdot&0  \\
\cdot&\cdot&\cdot&\cdot&\cdot&\cdot  \\
\cdot&\cdot&\cdot&\cdot&\cdot&\cdot  \\
0 &0 & 0& \Lambda \sqrt { \frac{n-1}{\kappa_{n-1}\kappa_{n-2} }}&2& \Lambda \sqrt { \frac{n}{\kappa_{n}\kappa_{n-1} }} \\
0 &0 &0 & 0& \Lambda \sqrt { \frac{n}{\kappa_n\kappa_{n-1} }}&2 \\
\end{array}
\right ) \nonumber
\end{eqnarray}
One can compute the determinant iteratively,
\begin{equation}
D_{m+1} (E,\Lambda)=2 D_m(E,\Lambda)- \Lambda ^2\frac{m+1 }{\kappa_{m+1}\kappa_m }D_{m-1}(E,\Lambda)\ .
\label{recursiond}
\end{equation}
The initial condition  $D_{-1}=1$ and  $D_0=2$  enables the subsequent computation of $D_m(E,\Lambda)$ for all $m \ge 1$.

 To simplify the notation we define $\eta = \frac{1}{2}-E$, and in the present chapter $\frac{1}{2}>\eta>0$. Also, the dependence on $\Lambda$ will be omitted. Thus,e.g.,  $ D_m(\eta) $ will stand for $D_m(E,\Lambda)$.

In the sequel we shall study in detail the recursion relation (\ref {recursiond}). Since
\begin{equation}
\hspace{-25mm}
\frac{m }{\kappa_{m-1}\kappa_m }=\frac{m }{\sqrt {(m-1+\eta)(m+\eta)}} =
1+\frac{\frac{1}{2}-\eta}{m} +{\mathcal O}(\frac{1}{m^2})  ,\
\label{estimate}
\end{equation}
in the  domain of  very large $m$ (\ref{recursiond}) reduces to
\begin{equation}
D_{m+1} (\eta)=2D_m(\eta)- \Lambda ^2 D_{m-1}(\eta)\ .
\label{recursionass}
\end{equation}
So that for large $m$, the $D_m$ are independent of $\eta$ and   $ D_m  \approx    \alpha (\xi_+)^m +\beta (\xi_-)^m $ where,
$$  \xi_{\pm} =  1 \pm \sqrt{1-\Lambda^2}\ . $$
The roots $\xi_{\pm}$ become complex beyond the critical value $\Lambda =1$. In the sub-critical domain ($1>\Lambda >0$ ), they are real  with $\xi_+> 1 > \xi_- >0$. For the rest of this section we shall restrict the discussion to the subcritical domain.

For intermediate values of $m$, and for $\Lambda \lesssim 1$ \,  the mild $m$ dependence of the term  $(1+\frac{\frac{1}{2}-\eta}{m})$ becomes significant, as can be expected from examining the local solution of the recursion relation in this  domain. That is, assuming a solution of the form $(\rho(m,\eta))^m$ where $\rho(m,\eta)$ is varying slowly with $m$, then,
\begin{equation}
 \rho_{\pm}(m,\eta)\approx 1 \pm \sqrt{1-(1+\frac{\frac{1}{2}-\eta}{m})\Lambda^2}\ .
\label{estim}
\end{equation}
Denote by $\tilde m_{t}(\eta) =\frac{\frac{1}{2}-\eta }{1-\Lambda^2}$ the largest value of $m$  for which $(1+\frac{\frac{1}{2}-\eta}{m})\Lambda^2
\ge 1$. Then, for\   $m\le \tilde m_{t}(\eta)$,\ \ $\rho_{\pm}(m,\eta)$ are  complex, while   for $m> \tilde m_{t}$,\ \ $\rho_{\pm}(m,\eta)$ are both real.   The oscillatory nature of the solution in the sub-critical  $m\le m_{t}(\eta)$ domain  depends sensitively on $\eta$ and hence the appearance of the zeros of the secular equations.
The heuristic arguments above, will be made more precise and quantitative in the next subsection.

In order to determine the correct asymptotic behavior for $m\gg m_{t}$, consider first the Jacobi matrix
$2{\bf I}+\tilde {\bf J_0}$ which is obtained from $2{\bf I}+{\bf J_0} $ (\ref{det}) by replacing the factors  $ \sqrt { \frac{n}{\kappa_n\kappa_{n-1} }}$ by $1$ throughout. The resulting finite determinants $\tilde D_m$ can be expressed in terms of $U_n(x)$ - the Chebyshev polynomial of the second kind,
\begin{equation}
\hspace{-10mm}
 \tilde D_n = \Lambda^{n}U_{n} (\frac{1}{\Lambda} )\ \overrightarrow { _{\ n\rightarrow \infty}}\ \ c(\Lambda)\xi_+^{n} , \ \ {\rm and} \ c(\Lambda) \ \ {\rm depends \ mildly \ on\ } \ \Lambda\ .
\label{chebas}
\end{equation}
$ J_0(\eta)-\tilde J_0 $ is  Hilbert-Schmidt, and therefore  $\frac{D(\eta)}{ \tilde D} $ exists. This regularization provides a bounded secular function  with zeros at the discrete spectrum of the original Hamiltonian for $|\Lambda|<1$ in the domain $0<\eta<\frac{1}{2}$.

The above discussion enables the introduction of two methods for computing the secular function. They are described in the following subsections.

\begin{figure}
	\includegraphics[width=.50\textwidth]{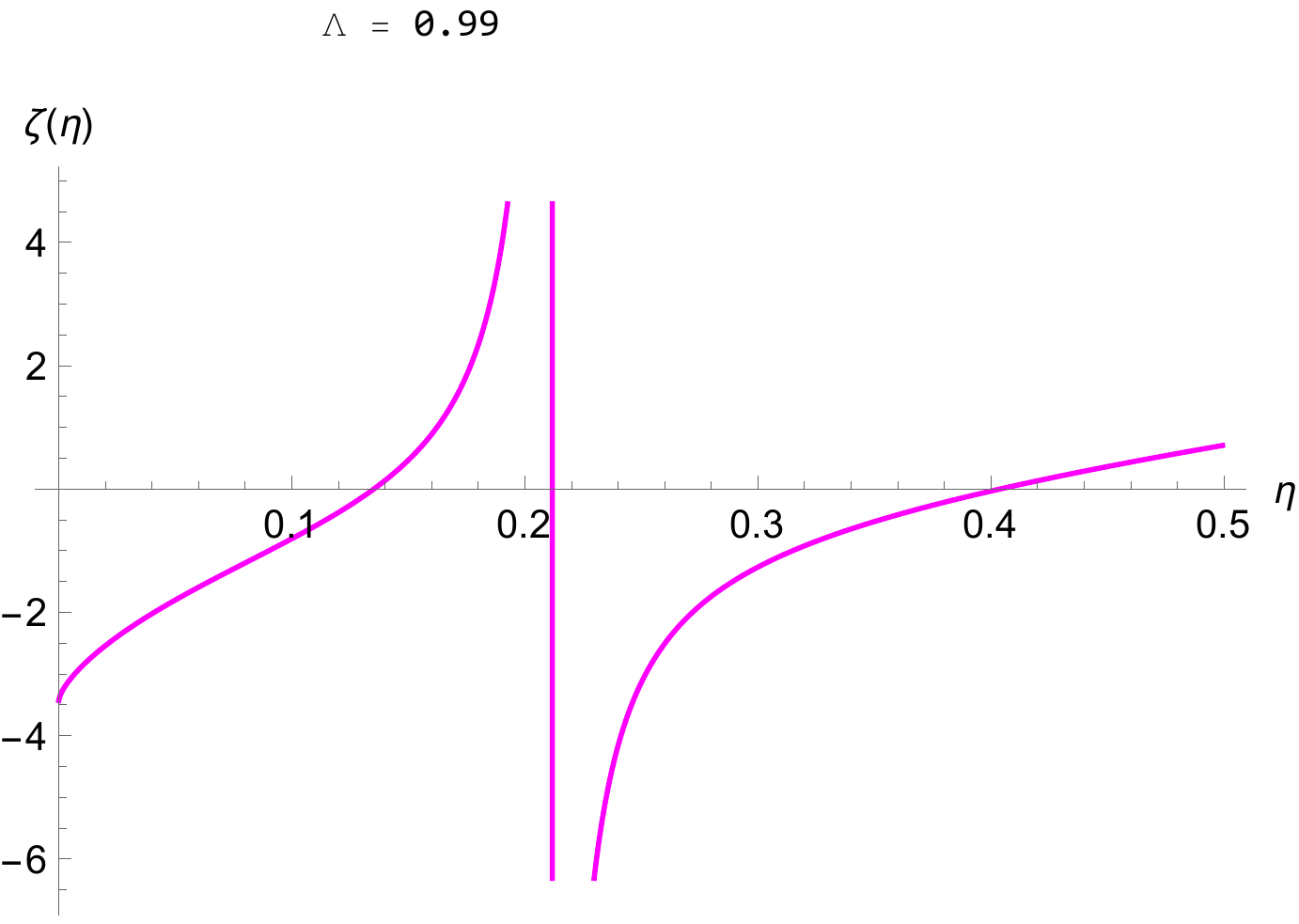}\hfill
	\includegraphics[width=.50\textwidth]{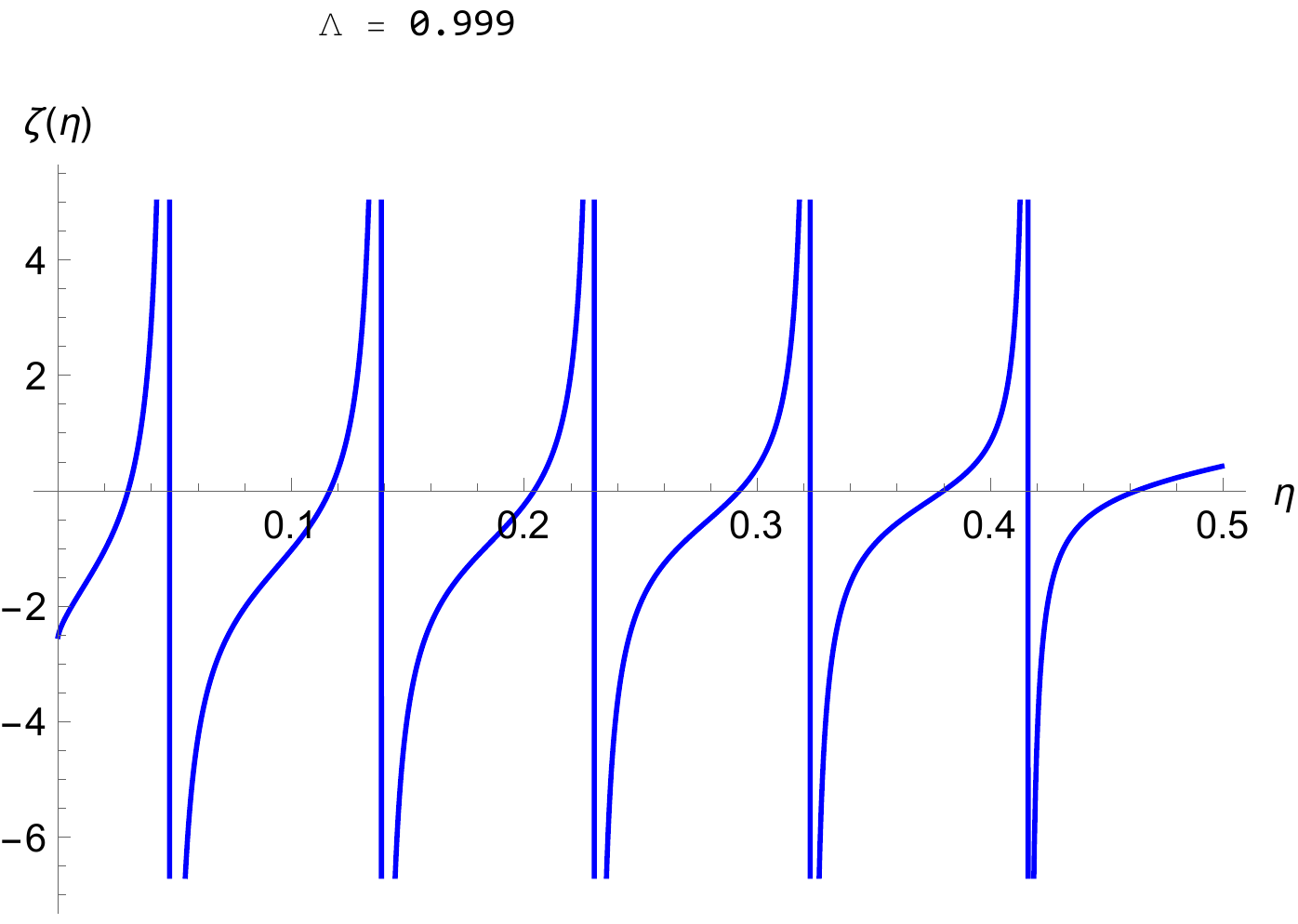}\hfill
\vspace{10mm}
	\includegraphics[width=.50\textwidth]{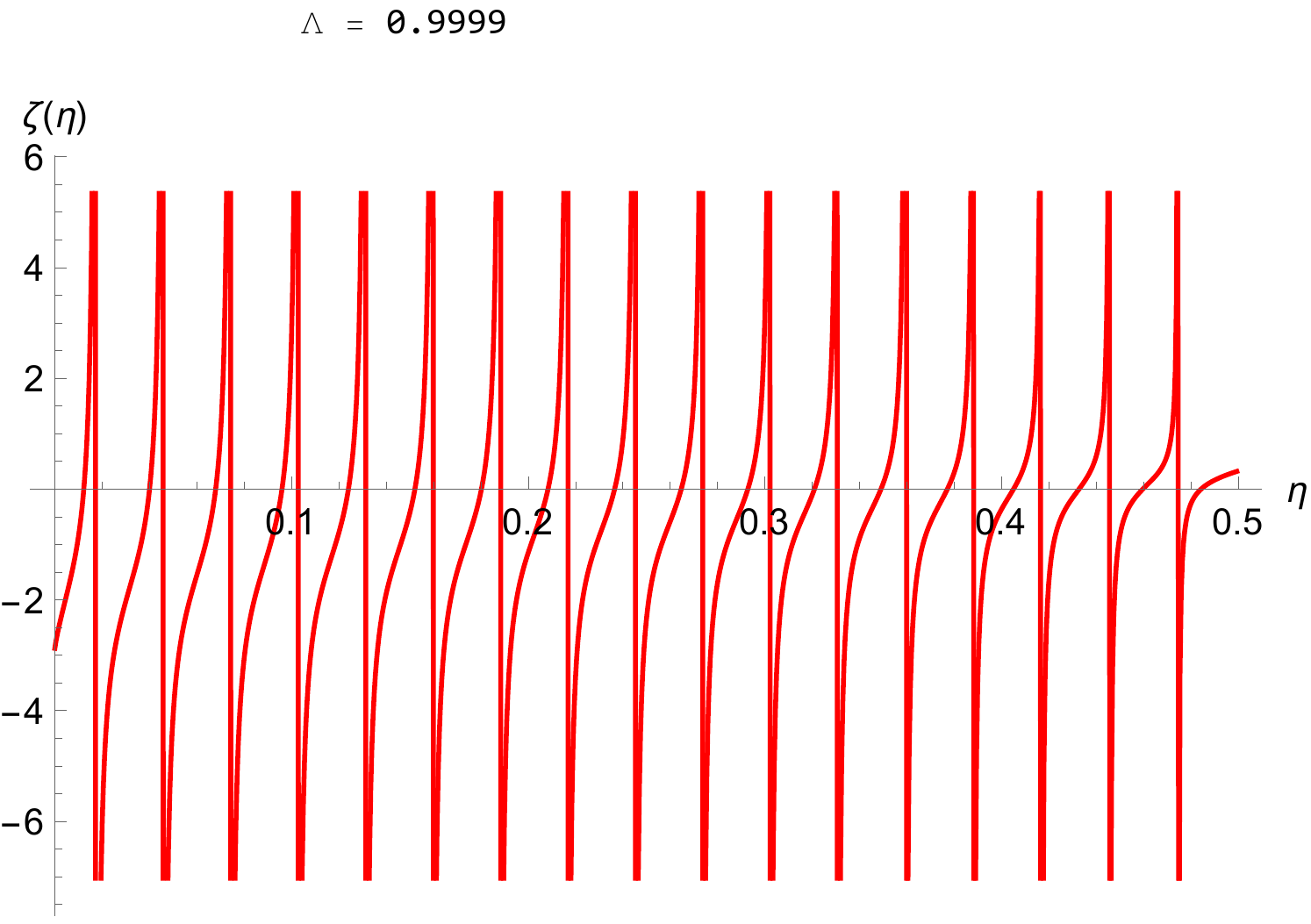}\hfill
	\includegraphics[width=.50\textwidth]{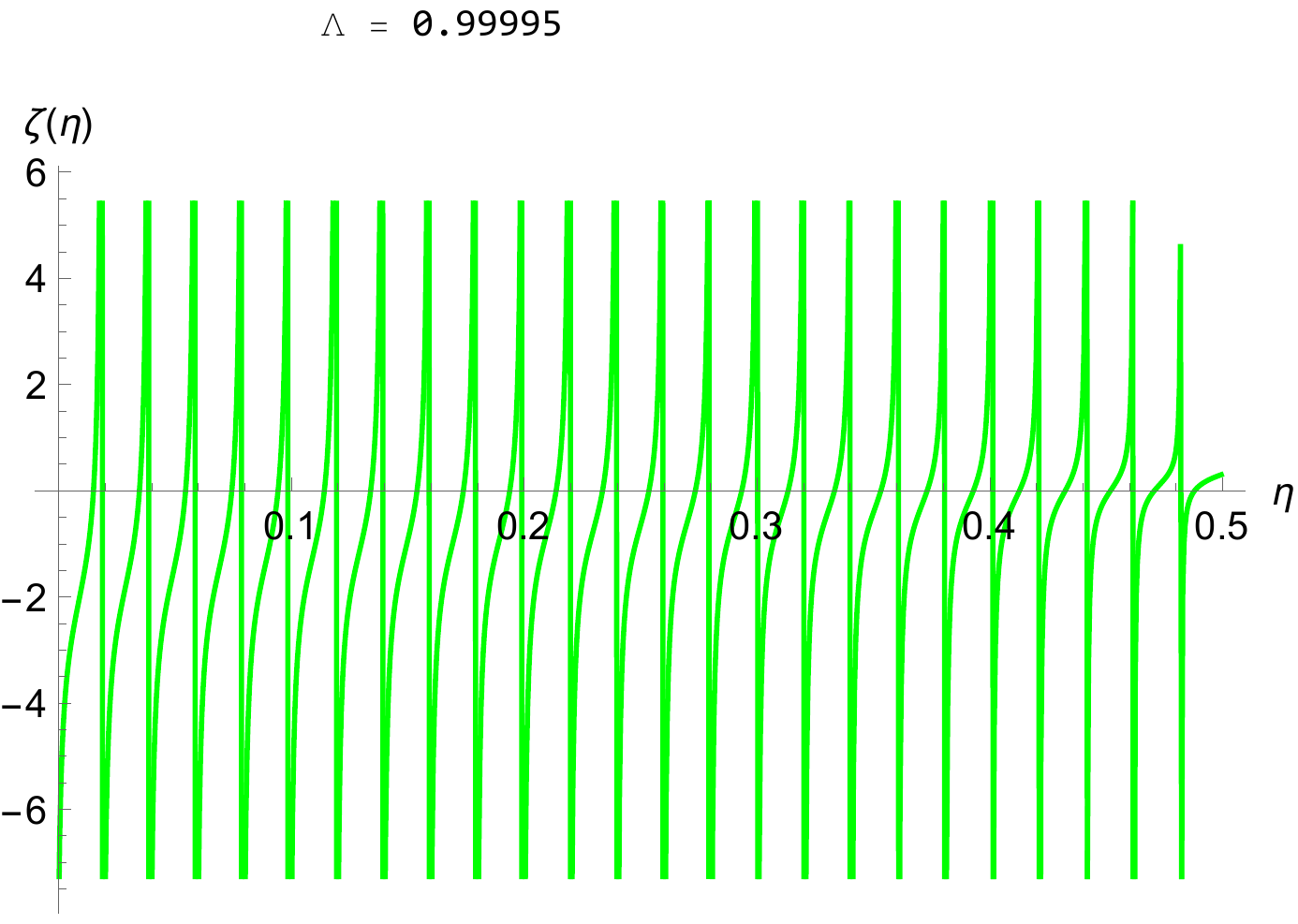}
	\caption{$\zeta_{I}(\eta)$, for a few values of $\Lambda$. The vertical lines are artifacts of the drawing program. The spectrum consists of the intersection of the smooth branches with the $\zeta=0$  axis. }
	\label{fig1}
\end{figure}
\subsubsection {A stable and fast iteration scheme:\hspace{80mm}  }
\label{subseciter}
 The regularization of the secular equation can be implemented by computing ratios of successive $D_m$ determinants.
 Denote $g_m(\eta) = D_{m}(\eta)/D_{m-1}(\eta)$. They satisfy
\begin{eqnarray}
\label{backg}
\hspace{-20mm}
{\it i} .& &\
g_{m}(\eta) = \Lambda ^2 \frac{m+1 }{\kappa_{m+1}\kappa_m }\frac{1}{2- g_{m+1}\ (\eta)} , ( {\rm   backwards\  recursion  }) \ . \nonumber \\
\hspace{-20mm} {\it ii} .& &\ g_1(\eta) =\frac{1}{2} (4-\frac{\Lambda^2}{ \kappa_0\kappa_1} ) \ \ ({\rm   to\  satisfy\  the\  first\  recursion\  line  })\nonumber \\
\hspace{-20mm} {\it iii} . & &\
g_m(\eta)\  \overrightarrow { _{\ m\rightarrow \infty}}=g_{\infty}=\xi_{+}= 1 +\sqrt{1-\Lambda^2}\ ,\ \ ({\rm   independent\  of\ \eta }) \   .
\end{eqnarray}
Starting with $g_{\infty}$ for sufficiently large $m$ and iterating the  backwards recursion (\ref{backg}.{\it i.}) results in  value for  $g_1(\eta)$. It is compared with the initial condition (\ref{backg}.{\it ii.}), resulting with the secular function
\begin{equation}
\zeta_{I}(\eta)= g_1(\eta)-\frac{1}{2} (4-\frac{\Lambda^2}{ \kappa_0\kappa_1}  )   ,
\end{equation}
its zeros provide the point spectrum.

Examples of $\zeta_{I}(\eta)$ computed for various values of the parameter $\Lambda$ as it approaches the critical value, are shown in figure (\ref{fig1}) .

\vspace {20mm}

\subsubsection {A Uniform WKB Approximation }
 \label{subsecwkb}
 \ \ \ \ \ \ \ \ \ \ \ \ \ \ \ \ \ \ \ \ \ \ \ \ \ \ \ \ \ \ \ \ \ \ \ \ \ \ \ \ \ \ \ \ \ \ \ \ \ \ \ \ \ \ \ \ \ \ \ \

\noindent The recursion relation (\ref{recursiond}) is approximated by
\hspace {-20mm}
\begin{eqnarray}
\hspace {-20mm}
D_{m+1} (\eta )-2 D_m(\eta)&=&-\Lambda ^2\frac{m+1}{\kappa_{m+1}\kappa_{m}}D_{m-1}(\eta)\\
\hspace {-20mm} &=&-\left [\left (\Lambda\sqrt{\frac{m+\frac{1}{2}}{m+\eta}}\right )\left
  (\Lambda\sqrt{\frac{m+1+\frac{1}{2}}{m+1+\eta}}\right ) +{\mathcal O}\left (\frac{1}{m^2}\right)
\right ]D_{m-1}(\eta)  . \nonumber
\end{eqnarray}
Write $D_m(\eta)= q_m(\eta)\ \prod_{r=0}^m \Lambda
\sqrt{ \frac{r+\frac{1}{2}}{r+\eta} }$. Then,
\begin{equation}
q_{m+1}+q_{m-1} -2\frac{1}{\Lambda}\sqrt{\frac{m+1+\eta}{m+1+\frac{1}{2}}}q_m=0
\end{equation}
were  $q_m$ stands for $q_m(\eta)$.  The equation above  takes a more suggestive form by writing it as
\begin{equation}
- ( q_{m+1}+q_{m-1}-2q_m)   -\left [2-  \frac{2}{\Lambda}\sqrt{\frac{m+1+\eta}{m+1+\frac{1}{2}}}\ \right ]q_m=0 \ .
\label{discrete}
\end{equation}
In this form it stands for a discrete version of a Schr\"odinger equation on the half-line, describing a particle of energy E=2, in a potential
\begin{equation}
\hspace{-15mm}
V(m)= \frac{2}{\Lambda} \sqrt{\frac{m+1+\eta}{m+1+\frac{1}{2}}} = \frac{2}{\Lambda} \sqrt{1-\frac{\frac{1}{2}-\eta}{m+\frac{3}{2}}}\  \approx  \ \frac{2}{\Lambda}  (1-\frac{1}{2}\ \frac{\frac{1}{2}-\eta}{\frac{3}{2}+m} )+ \mathcal{O}(\frac{1}{(m+\frac{3}{2})^2})\ .
\end{equation}
The potential is a smooth  and monotonic function of $m$,  increasing from
$V(0)=\frac{2}{\Lambda} \sqrt{\frac{1+\eta}{1+\frac{1}{2}}}$ to  $V(\infty)= \frac{2}{\Lambda} >2$. It is important to note that
the potential which depends on the two parameters $m$ and $\eta$, depends actually on a single scaled parameter   $\frac{\frac{1}{2}-\eta}{\frac{3}{2}+m} $. This fact will be used in the discussion at the end of the present section.
The value of $m$ for which $E=2=V(m)$ is the classical turning point  $m_t$:
\begin{equation}
m_t=\frac{\frac{1}{2}\Lambda^2-\eta}{1-\Lambda^2}-1 \ .
\label{turning}
\end{equation}
(Note: For $\Lambda^2 \lesssim 1$ the turning point $m_t$ is approximately equal to $\tilde m_t$ defined in the previous section).
Since $m_t>0$ is required to get a none trivial domain where $q(m)$ might cross zero, we get the condition
\begin{equation}
\eta_{max}= \frac{3}{2}\Lambda^2 -1> \eta > 0 \ \ \ {\rm and} \ \ \  \Lambda > \sqrt{\frac{2}{3}}\ .
\label{etamax}
\end{equation}

\begin{figure}	\includegraphics[width=.8\textwidth]{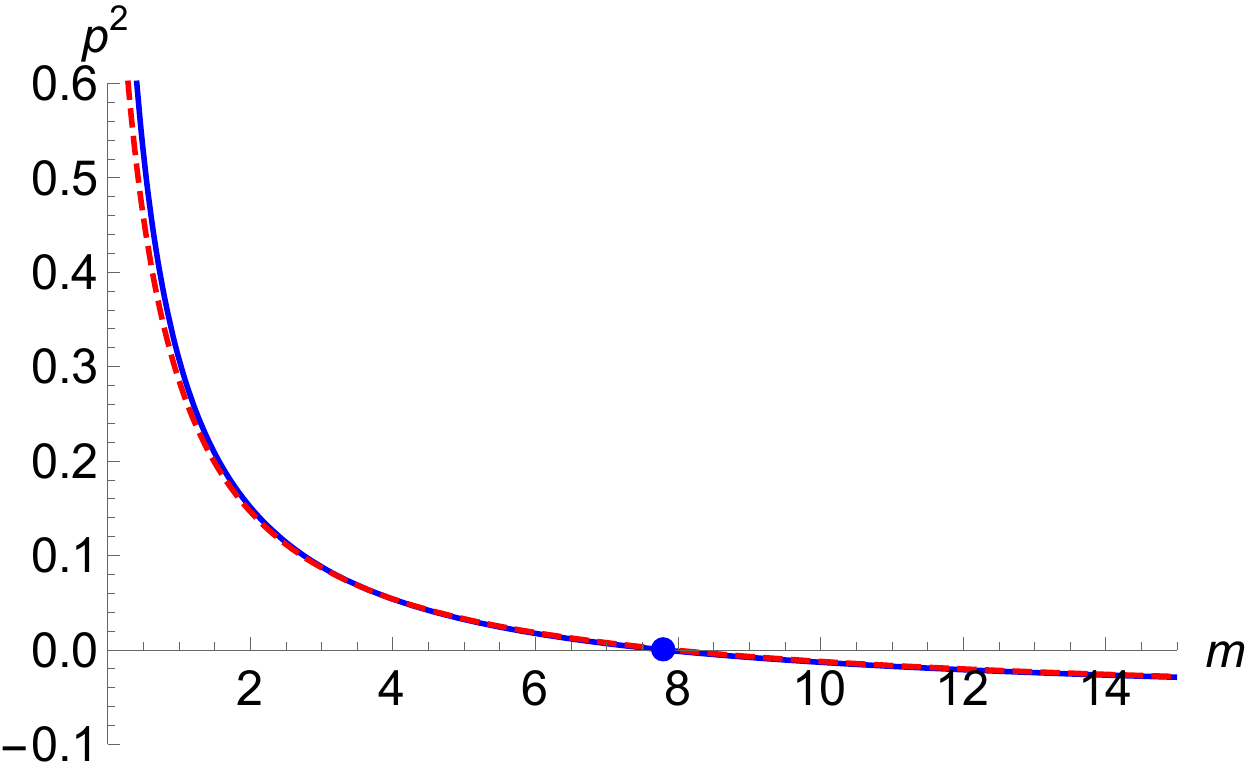}
		\caption{The local kinetic energy $p^2(m)$ (blue continuous line), and its leading $\frac{1}{m}$ approximation (red dashed line). The blue point marks the classical turning point. Data for   $\eta = .01 ,\ \ \Lambda =.97  $}
	\label{potential}
\end{figure}

The function $p^2(m)=(2-V(m))$ together with its leading approximation is shown in figure (\ref {potential}).

 The continuous version of (\ref{discrete}) is,
 \begin{equation}
 \label{wkb1}
-\frac{\partial^2 q(m,\eta)}{\partial m^2} -\left [2-  \frac{2}{\Lambda}\sqrt{\frac{m+1+ \eta}{m+1+ \frac{1}{2}}}\ \right ] q(m,\eta) = 0\ ,
 \ \  m \geq 1 .
\end{equation}
where $q(m,\eta)$ stands for a function of $m\in \mathbb{R^+}$ and assumes the values $q_m(\eta)$ for integer $m$. For the sake of notational simplicity, the parametric dependence of $q(m,\eta)$ on $\eta$ will be omitted until the discussion of the spectral secular equation.
To justify the transition from the discrete to the continuous Schr\"odinger operator, we use the fact that the potential term is analytic in the domain of interest, so that $q(m)$ can be expanded locally in a fast converging Taylor series. Using the notation  $q^{(n)}(m) = \frac{\partial^n  q(m)}{\partial m^n}$
we can then write
\begin{equation}
 q(m+1)-2q(m)+q(m-1) = q^{(2)}(m) + 2\sum_{n=2}^{\infty} \frac{ q^{(2n)}(m)}{(2n)!}
 \end{equation}
  Denote the classical local kinetic energy at $m$ by
 \begin{equation}
 p^2(m,\eta)=2 \left[1\ - \frac{1}{\Lambda}\sqrt{\frac{m+ 1+ \eta}{m+1+ \frac{1}{2}}}\right]\ ,
\end{equation}
Using $p^2(m)$ for short,  one can write the leading  approximation for the discrete equation as
 $$q^{(2)}(m)=-p^2(m) q(m)$$
 and use it to  expand successively the higher  terms in the Taylor series above. Thus,
\begin{equation} \hspace{-20mm}
q^{(4)}(m)=\frac{\partial ^2 q^{(2)}(m)}{\partial m^2}  = -\left[\left ( \frac{\partial ^2 p^2(m)}{\partial m^2}-(p^2(m))^2\right )q(m)+2\frac{\partial p^2(m)}{\partial m}\frac{\partial q(m)}{\partial m}\right].
\end{equation}
Locally, $|\frac{\partial q(m)}{\partial m}|\approx |p(m)||q(m)]$. Hence the absolute value of each term on the right hand side above can be written as a product of a function of $m$ times $|q(m)|$. Each of these function decreases at least as rapidly as $1/(m)^2$ for $m>5$. This follows from the fact that
\begin{equation}
p^2(m)\rightarrow 2[(1-\frac{1}{\Lambda}) -\frac{1}{2\Lambda} \frac{\frac{1}{2}-\eta}{m+\frac{3}{2}}] +\mathcal{O}\left(\frac{1}{(m+\frac{3}{2})^2}\right),
\end{equation}
and that the range of parameters of interest is $\Lambda \lessapprox 1$ and $\eta \gtrapprox 0$, so that the constant term in  $p^2(m)$ as well as the coefficient of  $\frac{1}{m+\frac{3}{2}}$ are small.

The smooth dependence of $V(m)$ on $m$ justifies the use of the WKB approximation in the present context.  Using the same arguments which were used to justify the transition from the discrete to the continuous Schr\"odinger operator, one can show that the criterion for the applicability of the WKB approximation is satisfied since  $\frac {|\frac{\partial p(m)}{\partial m} |}{p^2(m)}$ is of order $\frac{1}{m+1}$. Thus, the use of the continuous Schr\"odinger operator, and solving it using the WKB approximation are justified in the domain of interest here, provided that $ m> m_0$ and that at the same time $m_t$ is sufficiently larger than $m_0$. The turning point $m_t$  separates the real axis to two domains. The classically allowed domain is where $2>V(m)$  where $q(m)$ is oscillatory, whereas in the complementary, classically  forbidden domain, $q(m)$ is exponentially decaying or diverging. The primitive WKB approximation fails near $m_t$ and a uniform approximation which overcomes this difficulty is called for. The present case with a single classical turning point requires the use of a uniform approximation in terms of Airy functions \cite{berry}.

The WKB approximation for the  Schr\"odinger equation will be applied in the half-line $m \ge m_0$  and the condition on the wave function at the boundaries are the following :

\noindent \underline{The condition at $m \rightarrow \infty $} follows from the fact that the determinant $D_m$ must diverge exponentially with $m$ as was explained before. Therefore  in the asymptotic domain  $m \gg m_{t}$,\  $ q(m)$  is the exponentially increasing solution of (\ref{wkb1}).

\noindent \underline{The condition at $m=m_0$} is chosen such that one avoids the domain of $m$ values where neither the continuous version of the equation nor the WKB  approximations are sufficiently accurate. The boundary condition is derived by first iterating the exact discrete equations up to $D_{m_0+1}$ (\ref {recursiond}). Using the known factors which relate $D_{m}$ to $q(m)$, one requires
\begin{equation}
 \frac{q(m_0+1,\eta)}{q(m_0,\eta)} =\frac{D_{m_0+1}(\eta)}{D_{m_0}(\eta)}\frac{1}{\Lambda}\sqrt{\frac{m_0+1+\eta}{m_0+\frac{3}{2}}}.
 \label{bcairy}
\end{equation}

This requirement can be fulfilled only for a discrete set of $\eta$ which is the required spectrum. The spectral secular equation reads
\begin{equation}
\zeta_{II}(\eta) = \frac{q(m_0+1,\eta)}{q(m_0,\eta)}-\frac{D_{m_0+1}(\eta)}{D_{m_0}(\eta)}\frac{1}{\Lambda}\sqrt{\frac{m_0+1+\eta}{m_0+1+\frac{1}{2}}} =0.
 \label{secairy}
\end{equation}

The WKB approximation makes use of  the  classical action $s(m,\eta)$
\begin{equation}
 s(m,\eta) =   \int_{m_t}^{m} \sqrt {| (p^2(n,\eta))| }    {\rm d}n\ .
\end{equation}
The above integral can be evaluated, and it reads,
\begin{eqnarray}
\hspace{-22mm}
s(m,\eta) =
|p(m,\eta)|(m+\frac{3}{2}) -
\frac{\frac{1}{2}-\eta}{\sqrt{2 \Lambda }}
  [ \frac{ \arctan\hspace{-1mm}{\rm h} \left(  \frac{\sqrt {\Lambda}|p(m,\eta)| }{\sqrt {2( 1+\Lambda) }} \right)}{1+\Lambda}
       +\frac{      \arctan                \left(  \frac{\sqrt{\Lambda }|p(m,\eta)|}{\sqrt {2( 1-\Lambda) }} \right)}{1-\Lambda}
    ]
  \ .
\label{s-of-m}
\end{eqnarray}
  The explicit expression for $s(m,\eta)$  is used to define
 \begin{equation}
 \sigma(m,\eta) = {\rm Sign}[m-m_t]\ \frac{3}{2}\ \left [ s(m,\eta)\right ]^{\frac{2}{3}}\ .
\end{equation}
Then, the uniform WKB approximation for $q(m)$ which satisfies the boundary condition at large $m$ reads,
\begin{equation}
q(m,\eta)=
c \left[ \frac{\sigma(m,\eta)}{
  p^2(m,\eta) }\right]^{\frac{1}{4}}
{\rm Bi}(\sigma(m,\eta))\ ,
\label{q-airy}
\end{equation}
where $c$ is a normalization constant, and ${\rm Bi}(x)$ is the Airy function of the second kind. The boundary condition (\ref{bcairy}) is satisfied only for a discrete set of $\eta$ which forms the spectrum for the given value of $\Lambda$. This function can be now substituted in $\zeta_{II}$ and its zeros can be computed.

To get a better understanding of the spectrum, we shall use a simpler expression for $q(m,\eta)$ in the "classically allowed" domain $1\le m \le m_t(\eta)$. It is given by the asymptotic expression of $Bi(x)$ in the classically allowed region,
\begin{equation}
q(m,\eta) \approx c \frac{1}{\sqrt{|p(m,\eta)|}}\cos(s(m,\eta)+\frac{\pi}{4})\ .
\label{q-cos}
\end{equation}
With this approximation, and using the fact that $p(m_0,\eta$ is almost constant in the vicinity of $m_0$  the left hand side of (\ref{bcairy}) can be written as
\begin{equation}
\frac{q(m_0+1,\eta)}{q(m_0,\eta)}\approx 1+\frac{\partial q(m,\eta)}{\partial m}|_{m_0} \approx 1+\tan(s(m_0,\eta)+\frac{\pi}{4})\ p(m_0,\eta)\ .
\end{equation}
In the domain of interest, where $1-\Lambda$ is a small positive number, the dominant term in the action integral (\ref{s-of-m}) is
\begin{equation}
\hspace{-20mm}
\frac{\frac{1}{2}-\eta}{(1-\Lambda)\sqrt{2 \Lambda }}
   \frac{\frac{1}{2}-\eta}{(1-\Lambda)\sqrt{2 \Lambda }}
   \arctan [\mu (m_0,\eta,\Lambda)] \ , \ \rm{where} \ \ \mu(m_0,\eta,\Lambda) = \frac{\sqrt{\Lambda }|p(m_0,\eta)|}{\sqrt {2( 1-\Lambda) }} \ .
\end{equation}
Given $m_0$ and $\Lambda$, the range of $\eta$ is the interval $(0, \eta_{max}(m_0,\Lambda))$ where $\eta_{max}(m_0,\Lambda)$ is the value of $\eta$ for which $p(m_0,\eta) =0.$ For $(m_0+\frac{3}{2}) \ll (1-\Lambda^2)^{-1}$, $\eta_{max}$  approaches its maximum value $\frac{1}{2}$ so that the entire range of $\eta$ is almost completely covered. At the same time,
\begin{equation}
\mu= \frac{1}{  \sqrt{ 1-\Lambda}}\left(\frac{1}{2(m_0+\frac{3}{2})}-(1-\Lambda)\right )^{\frac{1}{2}},
\end{equation}
which grows indefinitely when $\Lambda$ approaches 1.  Thus, $\arctan [\mu (m_0,\eta,\Lambda)]$ starts at zero for $\eta=\eta_{max}$ and  immediately increases to its maximum value $\frac{\pi}{2}$. Therefore,
\begin{equation}
s(m_0,\eta) \approx  - \frac{\frac{1}{2}-\eta}{(1-\Lambda)\sqrt{2 \Lambda }}\frac{\pi}{2} \ .
\end{equation}
and the left hand side of (\ref{bcairy}) is
\begin{equation}
1-\tan[\frac{\frac{1}{2}-\eta}{(1-\Lambda)\sqrt{2 \Lambda }}\frac{\pi}{2}  +\frac{\pi}{4}]p(m_0,\eta) .
\end{equation}
This has poles inside every interval of length $ \pi $ of its argument. The right hand side of (\ref{bcairy}) is a smooth monotonic function of $\eta$ for small $m_0$. Hence, the number of eigenvalues in the interval $0 < E <\frac{1}{2}$ approaches
\begin{equation}
N(\Lambda) \approx  \left \lfloor \frac{{1}}{(1-\Lambda)2\sqrt{2 \Lambda }}\right \rfloor \ .
\end{equation}
 This result is consistent with the estimate given by  Solomyak and  Naboko. \cite{s2}

 \section{The Kronig-Penney model in a quadratic channel}
 \label{sectionkp}
 The periodic  Kronig-Penney model in a quadratic channel will be addressed here, using the scattering matrix developed in the preceding section, augmented by a  Floquet boundary  condition in the unit cell.  This provides the secular equation for the Floquet spectrum from which  the band spectrum of the periodic Shr\"odinger  operator (\ref {schroedinger}) is derived. Paper (I) in this  series \cite{Italo I} covers this subject using a different technique, and provides a detailed and broad view of the problem. Here, the scattering approach will be applied to a single case - the study of the band structure for $\Lambda<1$ and in the  domain $\frac{1}{2} >E>0$ .

 \label{second}
 Based on the Floquet theorem, the operator (\ref {schroedinger}) is  restricted to the unit cell
 \begin{equation}
 \hspace{-10mm}
 \label {schroedingerflo}
 H_{\omega}=
    -\frac{\partial^2\ }{\partial x^2} +\frac{1}{2}( -\frac{\partial^2\ }{\partial q^2} + q^2 ) +
 \lambda q\   \delta(x  ), \ \ \ {\rm for}\  | x |\le \frac{L}{2}-\epsilon \ \ \forall q
 \end{equation}
 together with  the boundary conditions
   that in the limit $\epsilon \rightarrow 0^+$, and for all $|\omega |\le \pi$
 \begin{eqnarray}
 \Psi (x=\frac{L}{2}-\epsilon,q)&=&
 e^{i\omega}\Psi (x=-\frac{L}{2}+\epsilon,q)
 \nonumber \\
 \frac{\partial \ }{\partial x }\Psi (x= \frac{L}{2}-\epsilon,q)   &=&e^{i\omega}
   \frac{\partial \ }{\partial x } \Psi (x=-\frac{L}{2}+\epsilon,q) .
 \label{bcflo}
 \end{eqnarray}
The  wave-function is expanded as in (\ref{wfunction})
\begin{eqnarray}
\Psi(x,q)&=& \Psi^{(a)}(x,q) \frac{1-{\rm Sign}[x]}{2}+ \Psi^{(b)}(x,q) \frac{1+{\rm Sign}[x]}{2}, \ |x|\le \frac{L}{2} \nonumber \\
\Psi^{(a)}(x,q) &=& \sum_{m=1}^{\infty}  \frac{1}{\sqrt{| k_m|}}(a_m^+ e^{+ik_mx} +a_n^- e^{-ik_mx} ) f_m(q) \nonumber \\
\Psi^{(b)}(x,q) &=& \sum_{m=1}^{\infty}  \frac{1}{\sqrt{| k_m|}}(b_m^+ e^{+ik_mx} +b_m^- e^{-ik_mx} )f_m(q) ,
\label{wfunction1}
 \end{eqnarray}

 Following the method of paper (I) and repeating it for the sake of completeness,  one can make use of the symmetry of the Schr\"odinger operator in the unit cell   under reflection about $x=0$ followed by conjugation. Any solution should have this symmetry, which implies that
 \begin {equation}
 \hspace {-23mm}
\int_{-\infty}^{\infty} f_n(q)  \Psi (x,q ){\rm d}q =r_n \left (\sin [k_n(L-|x|)] +e^{i\omega \frac{x}{|x|}} \sin [ k_n | x | ] \right ) \     ;    \ x \in {\mathbb R},\   \forall n\in{\mathbb N}^0
 \label{symwfho}
 \end {equation}
 where $r_n$ are real.
 The expression above can be written separately for the positive and negative values of $x$, and then compared with (\ref {wfunction1}) to get a relation between the coefficients $a_n^{\pm},b_n^{\pm}$ on the one hand and the $r_n$ and $\omega$ on the other hand. Using the vector notation introduced in the previous section, these relations read
 \begin{equation}
{\bf a}^{\pm}=\pm\ \frac{1}{2i}\ {\bf r}\ (e^{\pm i{\bf K}L}-e^{-i\omega}) \   \ ;
  \ \  {\bf b}^{\pm}=\mp\ \frac{1}{2i}\ {\bf r}\ (e^{\mp i{\bf K}L}-e^{i\omega})\ ,
\label{italo}
 \end{equation}	
where ${\bf K  }$ is a diagonal matrix with entries  $ \{k_n\}_ { n \in {\mathbb N}^0 }$.

 The effect of the $\delta$ interaction at $x=0$ is expressed by requiring that the incoming and outgoing amplitudes are related by the scattering matrix
 \begin{eqnarray}
 \left ( \begin {array} {l}
 {\bf a}^{-}\\
 {\bf b}^{+}\\
 \end{array}\right )=
 {\bf S}(k)
 \left ( \begin {array} {l}
 {\bf a}^{+}\\
 {\bf b}^{-}\\
 \end{array}\right )\  ,
 \label{rearanger1}
 \end{eqnarray}
 with
  (\ref{smatrixq})
 \begin{eqnarray}
 {\bf S}(k)=- \left (
 \begin {array} {cc}
 {\bf I} &0 \\
 0 &{\bf I}\\ \end{array}\right )
 +\left [{\bf I} +\frac{1}{2}{\bf J_{0}}(k) \right ]^{-1}
 \left (  \begin {array} {cc}
 {\bf I} &{\bf I} \\
 {\bf I} &{\bf I}\\ \end{array}\right )\ .
 \label{smatrixq1}
 \end{eqnarray}
  Substituting (\ref{italo}) in (\ref{rearanger1}) one gets that the vector ${\bf r }$ should satisfy a homogeneus linear equation, which has a non-trivial solution only if the determinant of the matrix involved vanishes. One obtains the condition
 \begin{equation}
 \det \left[ (2{\bf I} +{\bf J_0 }) \sin {\bf K}L  + 2 i (e^{i{\bf K}L}-{\bf I}\cos \omega ) \right ] =0
 \end{equation}
 which is the secular equation providing for each value of $\omega$ the set of $k_n(\omega)$ - the Floquet spectra which form the spectral bands if they exist. In particular, we focus on  the  bands in the energy range $0\le E \le \frac{1}{2} $ by setting $i\kappa_n$ in ${\bf K}$ to obtain
  \begin{equation}
  \det \left[  2(\cosh{\bf K }L -{\bf I} \cos \omega)   + {\bf J_0 } \sinh {\bf K }L  \right ] =0.
 \label{secularkph0}
 \end{equation}
 The secular function is deduced after regularizing the above determinant,
 \begin{equation}\
\zeta (\eta, \omega)=\det \left[  2({\bf I} -  \cos \omega (\cosh{\bf K }L )^{-1}) + {\bf J_0 } \tanh {\bf K }L  \right ] =0.
\label{secularkph}
 \end{equation}
 Let $d_n(\eta,\Lambda,\omega)$ with $n\in \mathbb{N}_{0}$ be the upper main minor of dimension $(n+1)\times (n+1)$ of the infinite matrix whose determinant is the $\zeta$ function defined above. It is the generalization of the matrix which appears in (\ref{det}) (with $E=\frac{1}{2}-\eta$).
 \begin{equation}
 \hspace{-25mm}
 (d_n(\eta,\Lambda,\omega,L))_{m,m'}=  \delta_{m,m'}\ 2 \left (1- \frac{\cos \omega }{\cosh\kappa_mL} \right ) + ({\bf J_0})_{m,m'} \tanh\kappa_{m'}L , \ \ \ 0 \le  m,m' \le n .
 \end{equation}
The determinants $D_n(\eta,\Lambda,\omega,L)= \det[d_n(\eta,\Lambda,\omega,L)]$ satisfy a similar recursion relation as in (\ref{recursiond}) albeit with modified coefficients as follows from the definition of $(d_n(\eta,\Lambda,\omega,L))_{m,m'}$:
\begin{equation}
D_{m+1} = v_{m+1}D_m -\Lambda^2 \frac{m+1}{\sqrt{(m+\eta)(m+1+\eta)}}u_m u_{m+1} D_{m-1}
\label{kprecursion}
\end{equation}
where
\begin{equation}
v_m= 2(1-\frac{\cos\omega}{\cosh ( \sqrt{ m+\eta}\ L) })\ \ {\rm and} \ \ u_m =\tanh( \sqrt{ m+\eta}\ L)\ .
\label{kpdefuv}
\end{equation}
Clearly, both the $|v_m|$ and the $|u_m|$ are bounded from above for all values of the parameters $\omega$ and $\eta$.  When $m$ increases,  $v_m \rightarrow 2 + \mathcal{O}(e^{-\sqrt{m}L})$ and $u_m\rightarrow 1+ \mathcal{O}(e^{-\sqrt{m}L})$. Therefore the asymptotic behavior of the solutions of (\ref{kprecursion}) coincide with those of (\ref{recursiond}). The initial conditions for the recursion are
\begin{equation}
 D_0= v_0 \ \ {\rm and} \ \   D_1=v_0 v_1-\frac{\Lambda^2}{ \sqrt {\eta(1+\eta)}}\ u_0 u_1.
 \label{kpbcond}
\end{equation}

\begin{figure}
 	\includegraphics[width=1.0\textwidth]{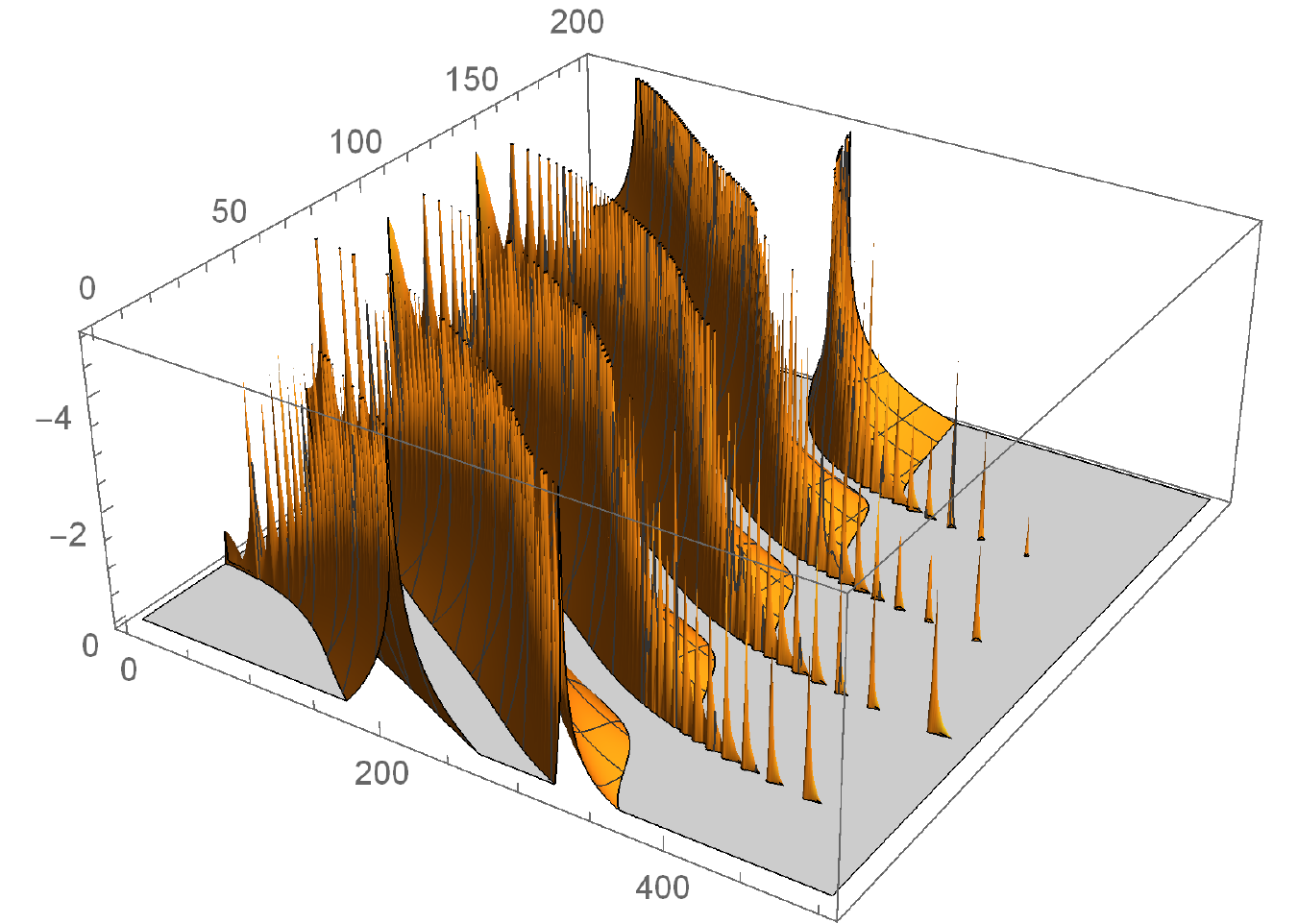}
 	\caption{A three dimensional view of $\log[|\zeta(\eta,\omega)|]$ restricted to negative values. Here $L=1.75$ and $\Lambda=.999$.
 A rectangular grid of $200 \times 500$ points is used to represent $\log[|\zeta(\eta,\omega)|]$ in the $(\eta,\omega)$ plane, with $200$ points in the interval $0\le \eta\ \le .5$, and $500$ in the interval $0\le\omega\le \pi$.  The spikes indicate zeros or poles of the function. (The distribution of zeros for this case is shown also in the first frame in the middle line of Figure (\ref{figbands}).}
 	\label{kpb175-3d}
 \end{figure}

\begin{figure}
 	\includegraphics[width=.40\textwidth]{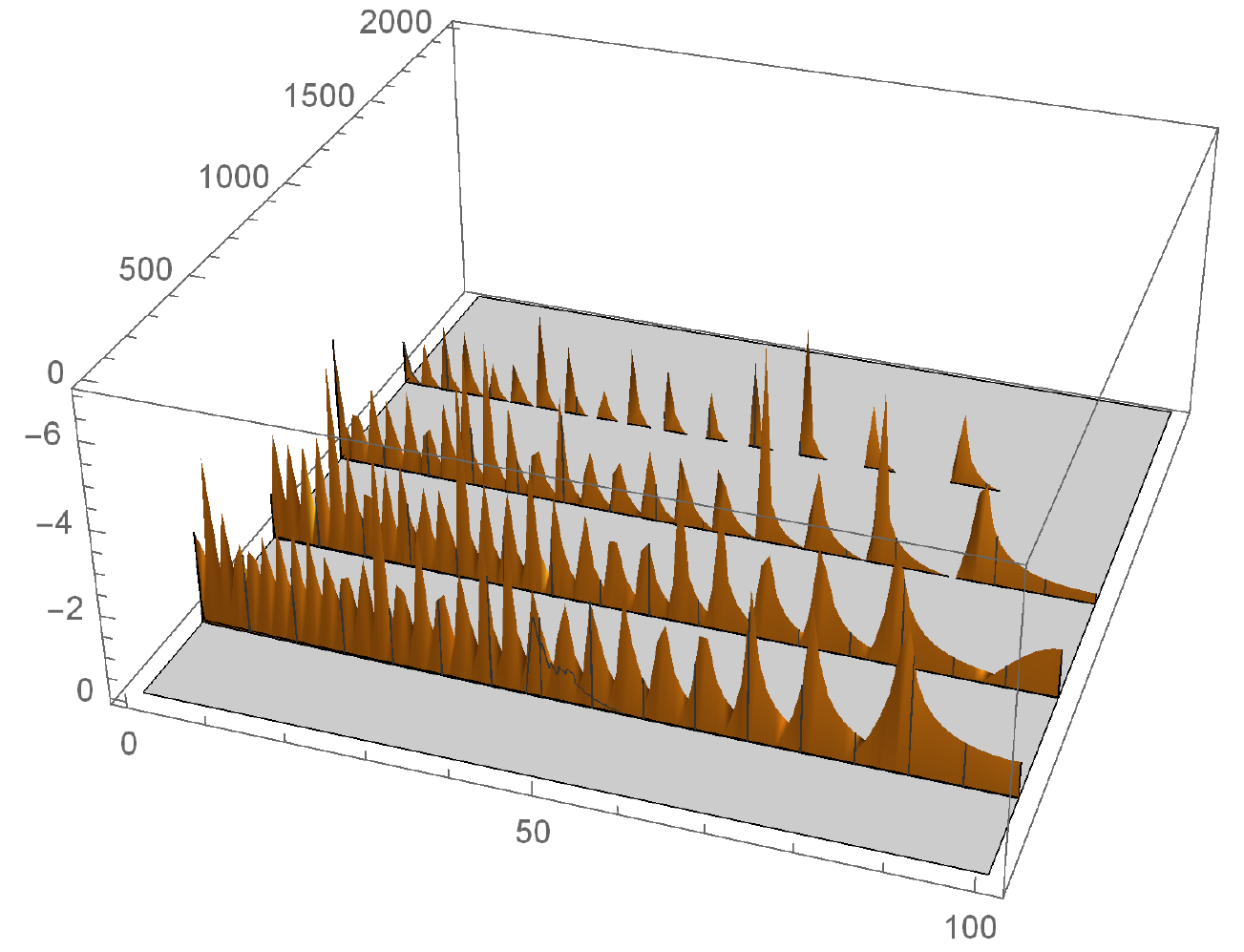}\hfill
    \includegraphics[width=.40\textwidth]{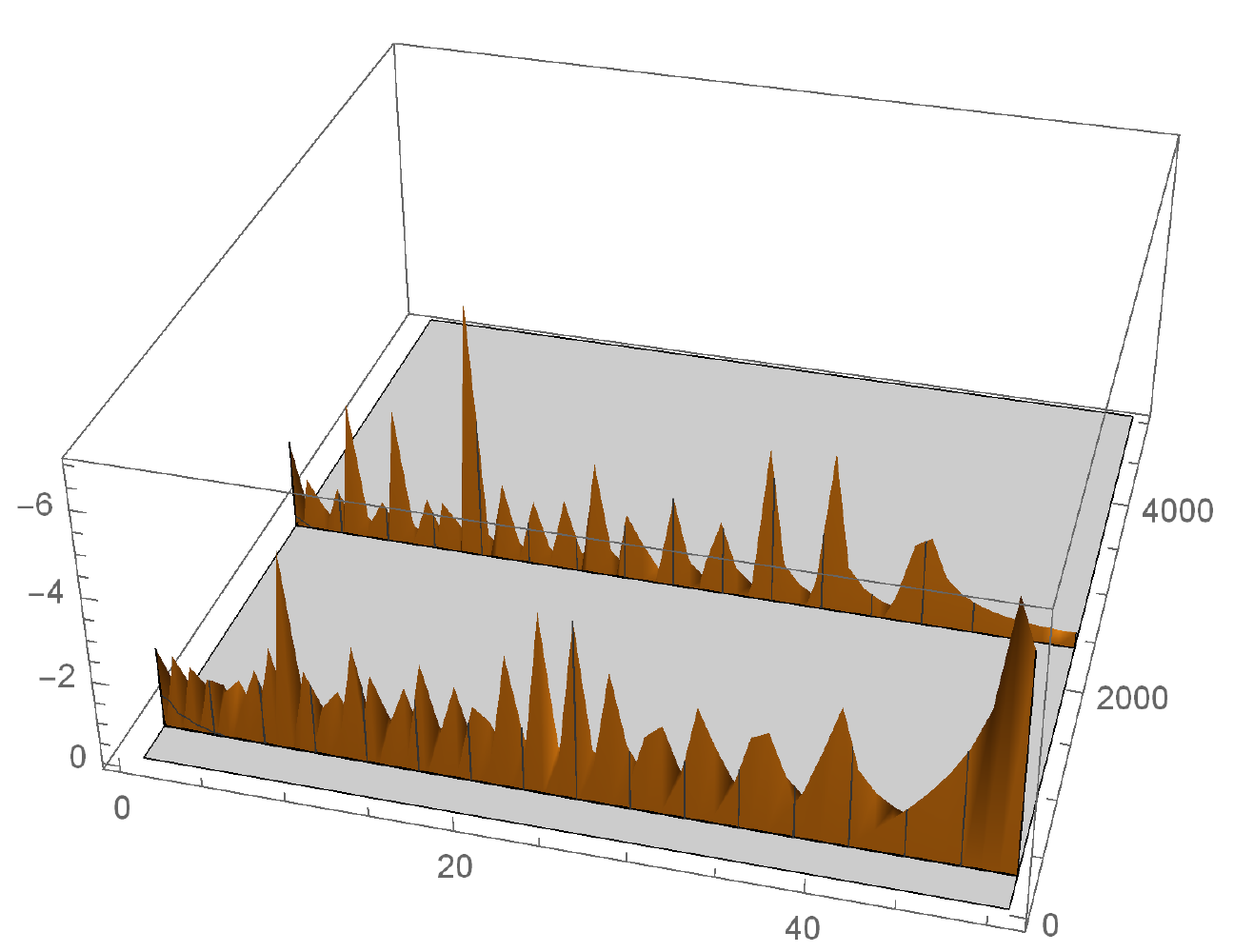}\hfill
 	\includegraphics[width=.40\textwidth]{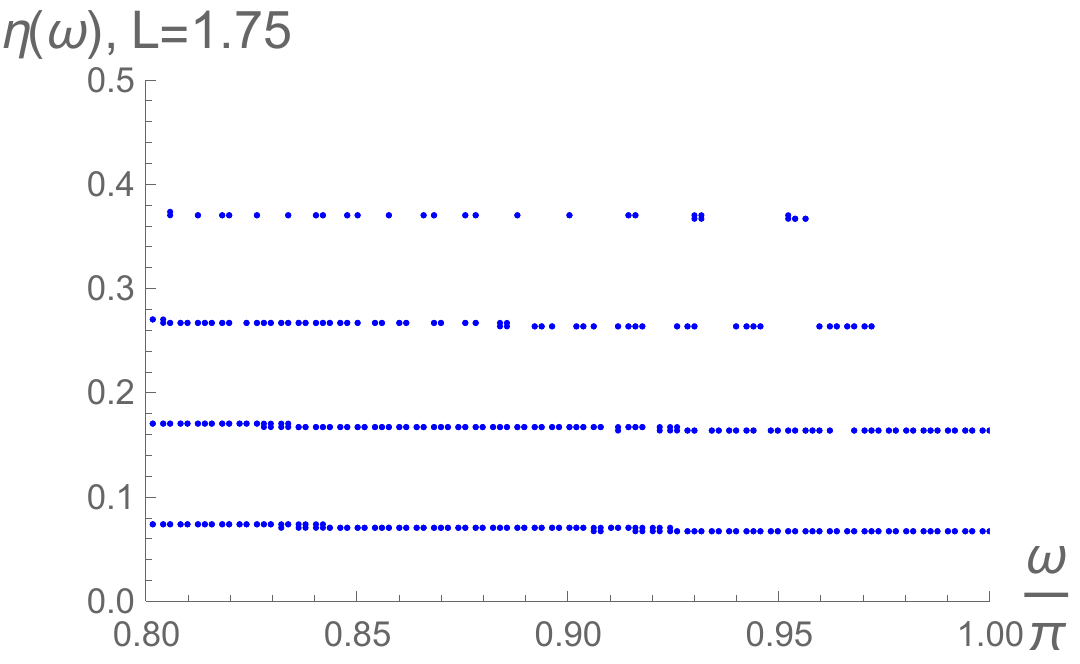}\hfill
    \includegraphics[width=.40\textwidth]{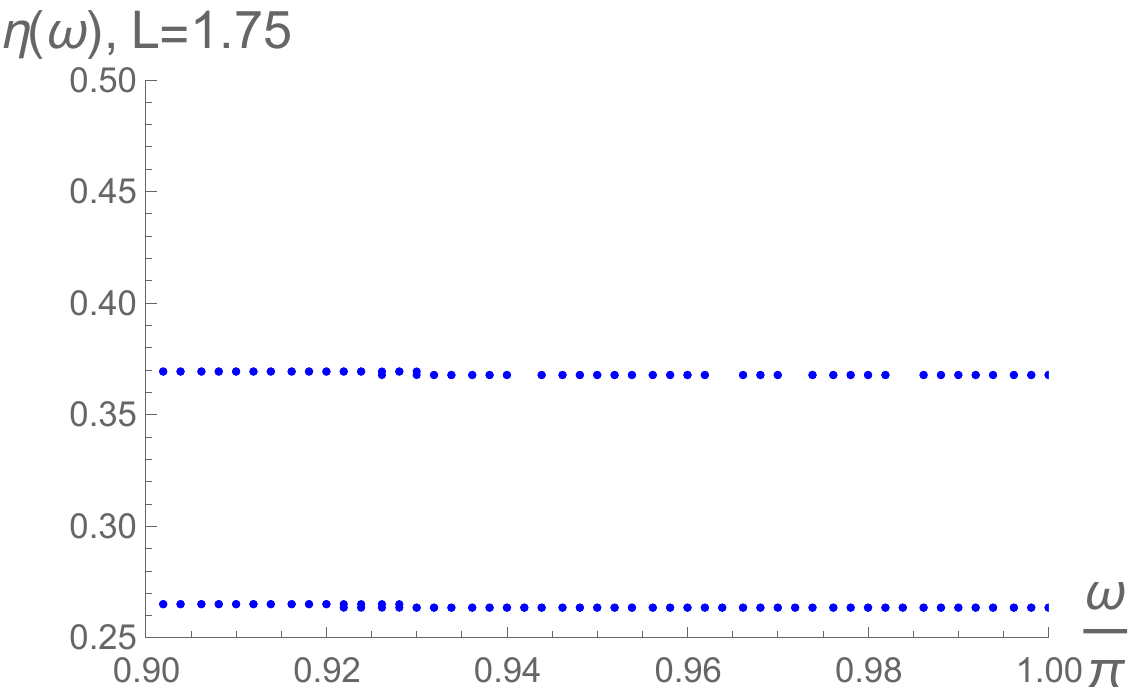}\hfill
 	\caption{Higher resolution for $\Lambda=.999, \ \ L=1.75$. The step size in the $\omega$ variable is the same as in the previous figure. The $\eta$ step-size is decreased by a factor $.1$ (left) and another $.1$ (right). Note: The two Floquet spectra  on the right frames are the ones appearing as the top curves on the left frames.}
 	\label{highres}
 \end{figure}

With this information, one can adopt the method described in  subsection \ref {subseciter}  to  compute the spectrum for every value of $\omega$. In the following
paragraphs the numerical computation of the Floquet spectrum will be described. The starting point is the computation of the secular equation \ref{secularkph}. It was computed by a backward iteration with the appropriate initial conditions and once reaching $m=1$ the $\eta$ spectrum is identified as the $\eta$ values for which the boundary conditions are satisfied. The secular equation  was computed on a $500\times 200$ grid of points  on the rectangular domain $(0 \le  \omega \le \pi )\times  (0\le \eta \le 1/2)$. More dense grids were also used for checking purposes.
Figure (\ref {kpb175-3d})  shows the numerical values of   $\log[|\zeta(\eta,\omega)]$ computed on the grid mentioned above for $L=1.75$ and $\Lambda=.999$. Restricting the view to negative values only, the high spikes correspond the points where the function approximately vanishes, which is the domain where the zeros of the $\zeta$ function are to be located.  The  spikes are concentrated along lines and in some part of the $(\eta,\omega)$ plane they are quite dense. In other parts, the density of spikes thin out to isolated sections or points. This occurs in domains  where the absolute values of $\log[|\zeta(\eta,\omega)]$ are appreciably reduced.  Further tests with denser  grids show more spikes  which interpolate between the ones found with the lower resolution. This is demonstrated in Figure (\ref{highres}) where increasing the $\eta$ resolution in two steps, while keeping the $\omega$ step-size constant show how the spike bands are completed in the $\zeta$ functions (upper frames)  and that more  points fill the gaps in the  lines which represent the Floquet spectra (lower frames).

The zeros  were identified on the lines of constant $\omega$ by locating neighboring $\eta$  points  where the secular function change signs, and using linear interpolation to locate the zero. Further tests were applied to exclude spurious zeros.

\begin{figure}
 	\includegraphics[width=.33\textwidth]{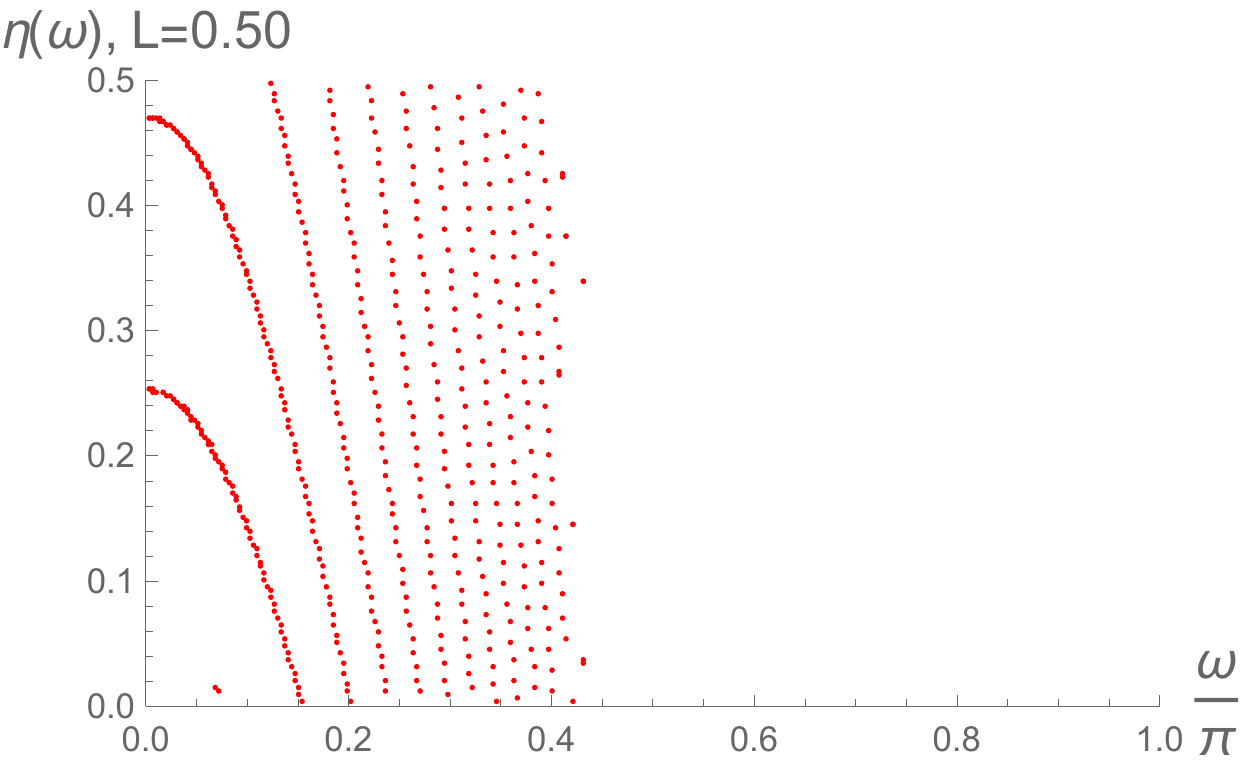}\hfill
     \includegraphics[width=.33\textwidth]{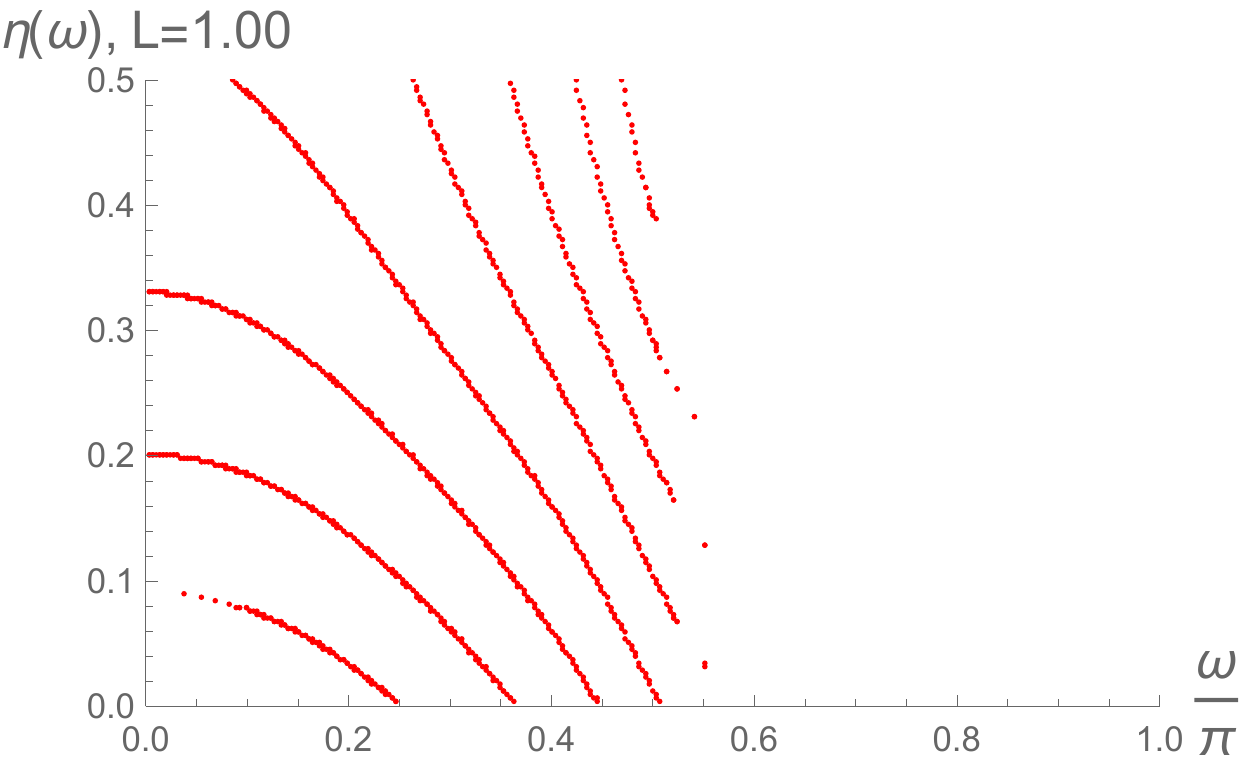}\hfill
 	\includegraphics[width=.33\textwidth]{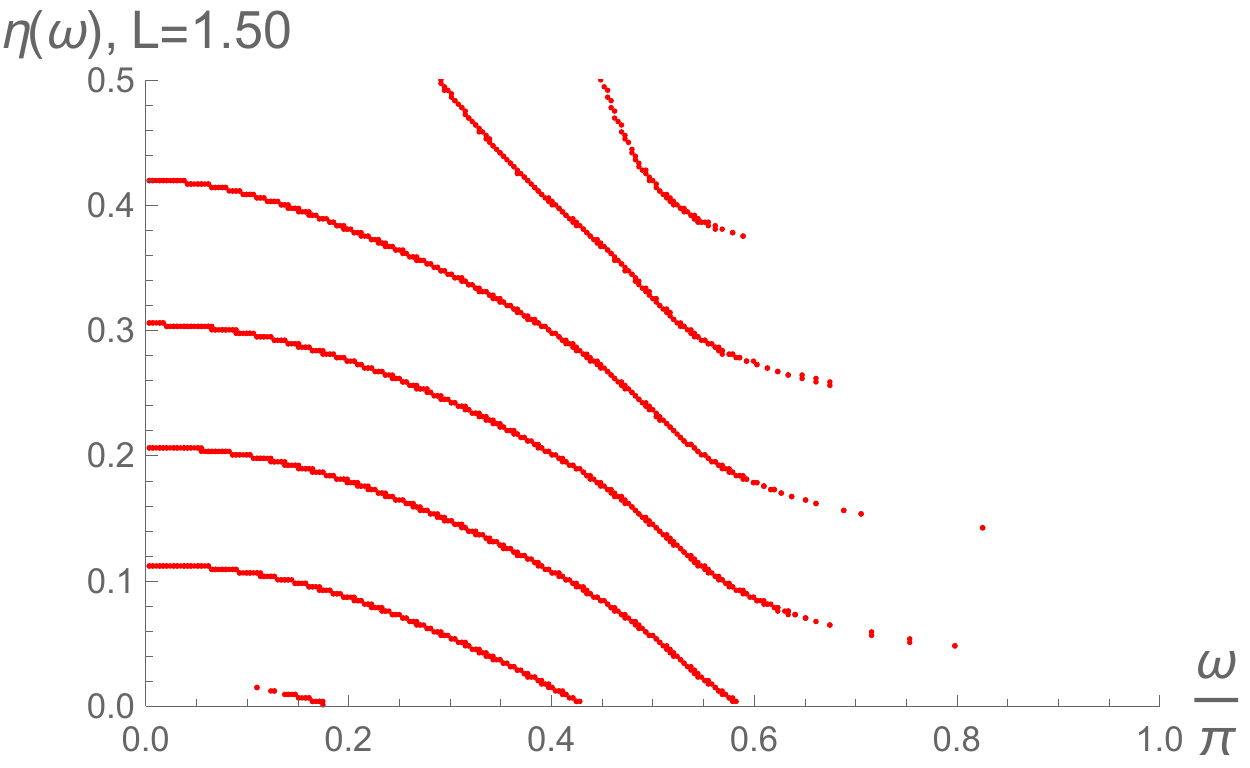}\hfill
    \includegraphics[width=.33\textwidth]{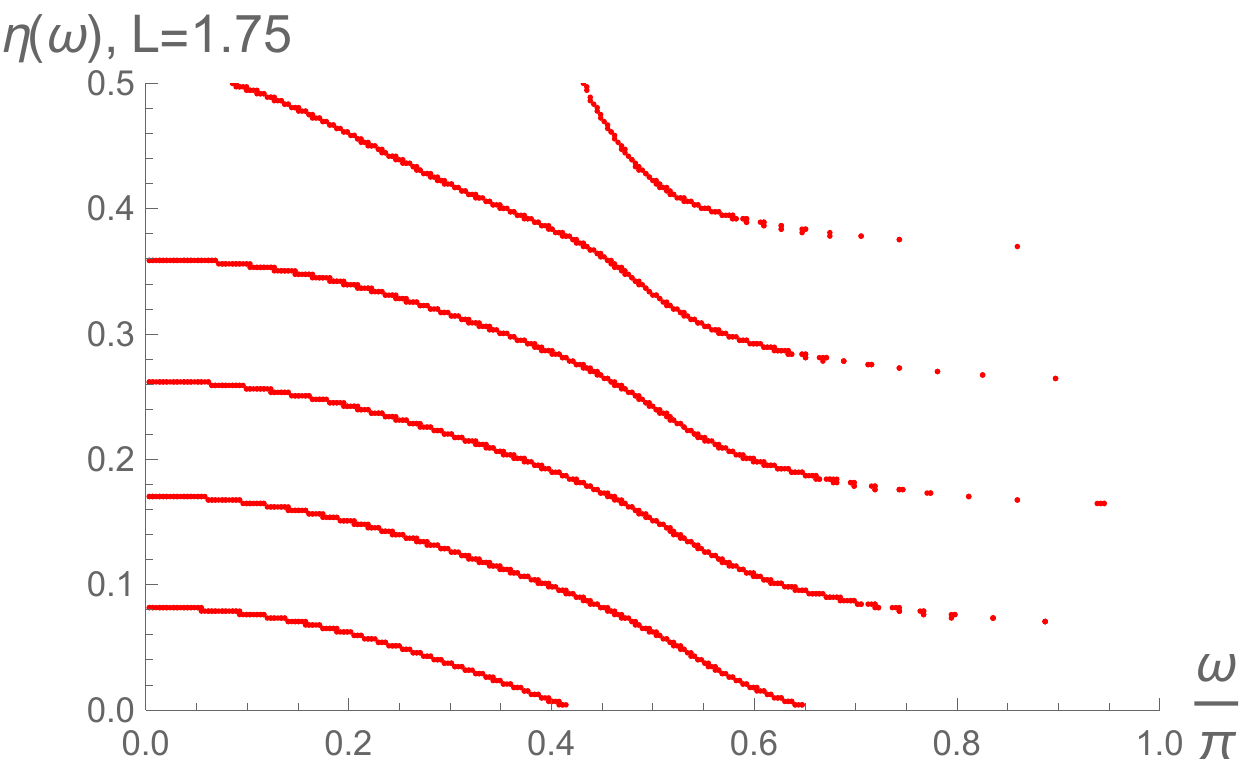}\hfill
 	\includegraphics[width=.33\textwidth]{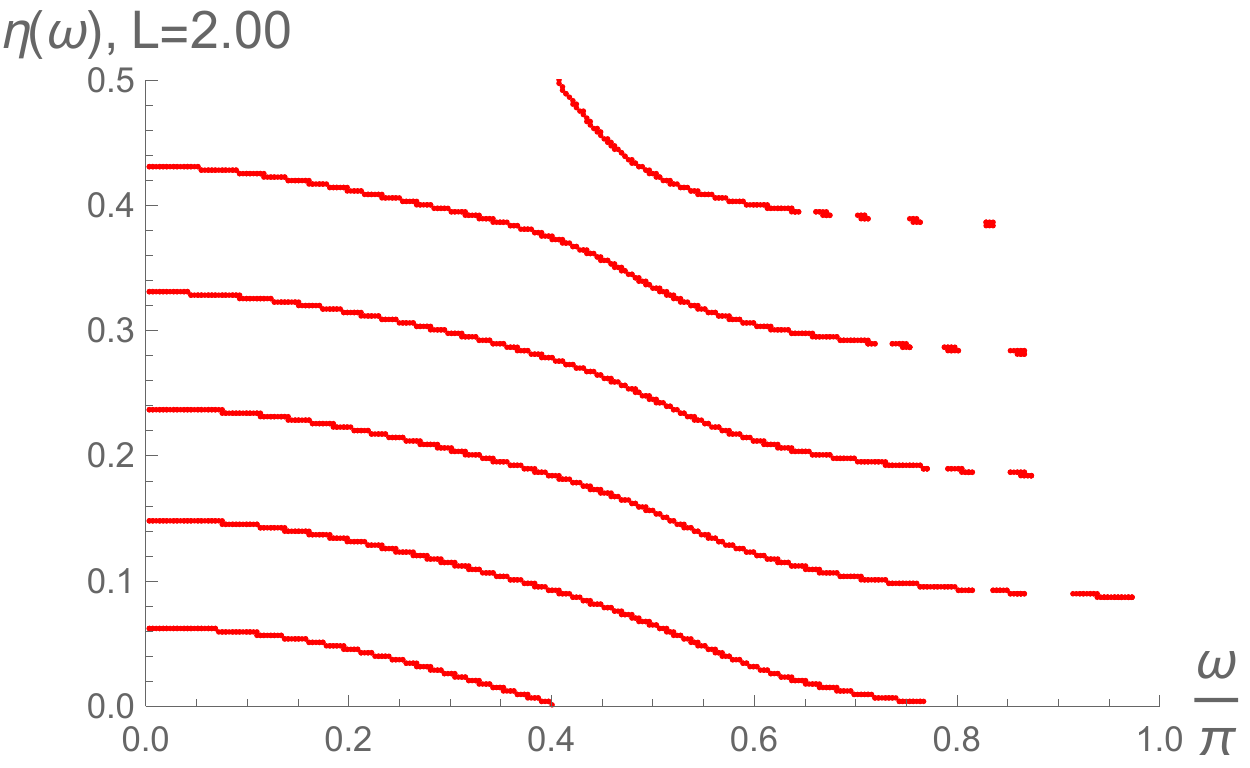}\hfill
    \includegraphics[width=.33\textwidth]{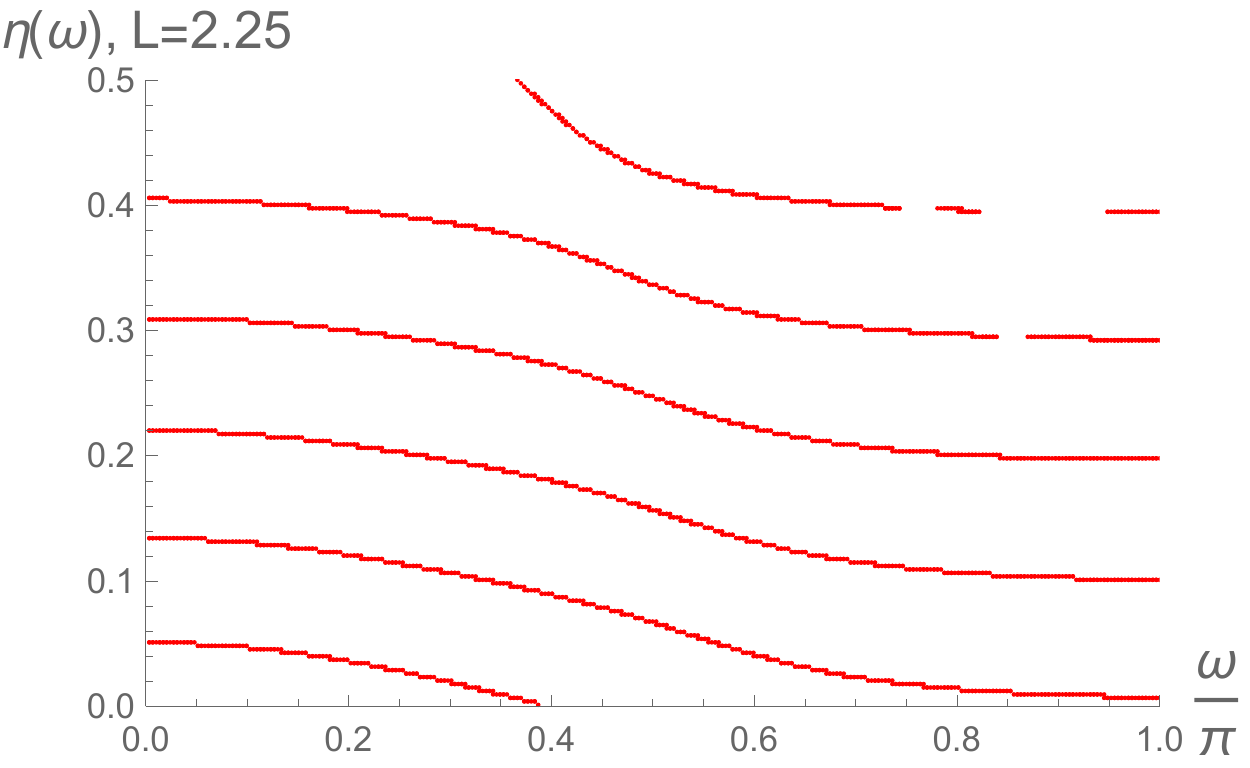}\hfill
 	\includegraphics[width=.33\textwidth]{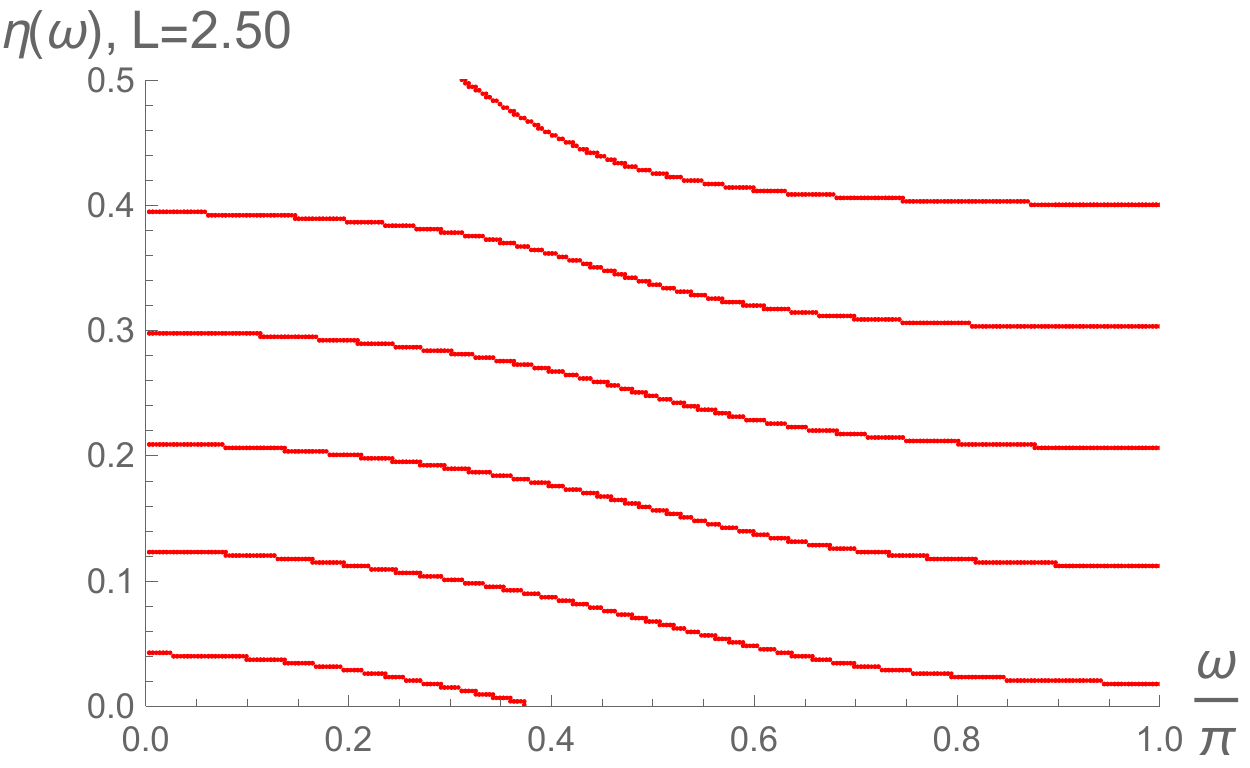}\hfill
    \includegraphics[width=.33\textwidth]{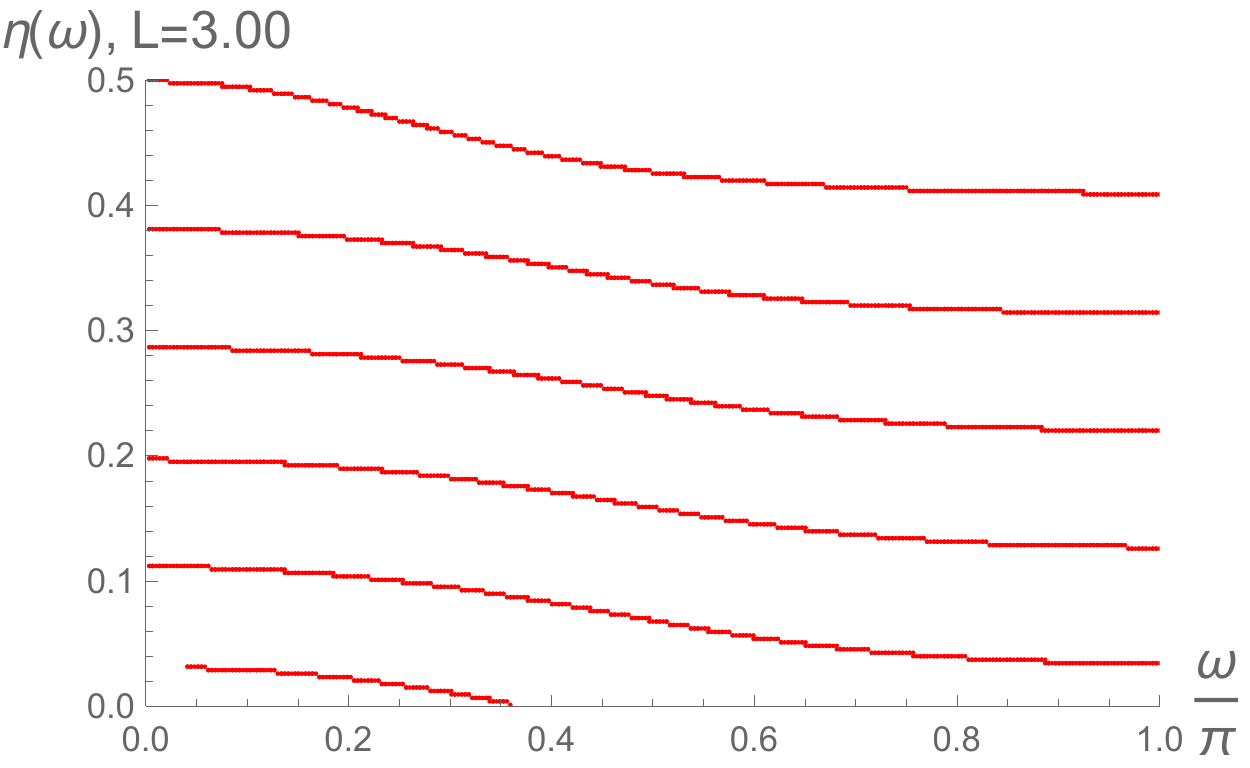}\hfill
    \includegraphics[width=.33\textwidth]{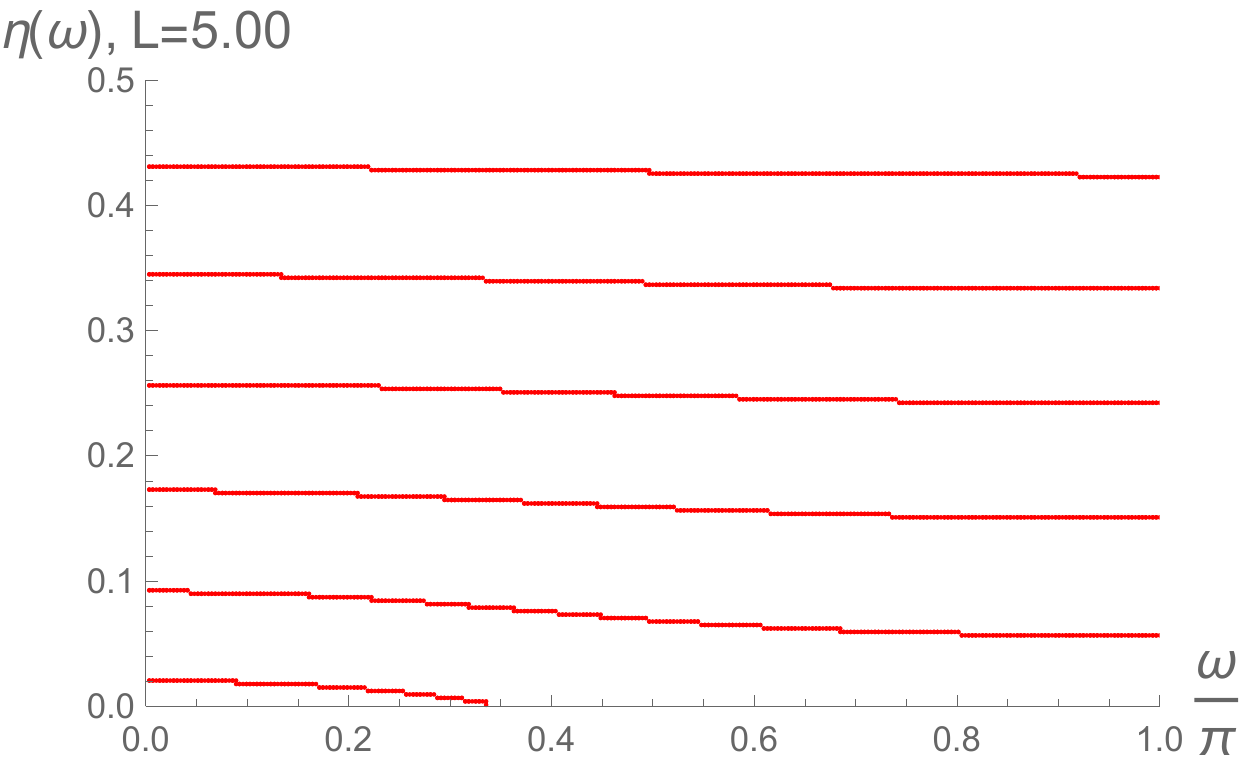}
 	\caption{The Floquet spectrum $\eta (\omega,L)$,  for  $\Lambda =.999$ and increasing values of $L$ from top left to bottom right: $L= .5,1.,1.5,1.75,2.,2.25,2.5,3.,5.$ .}
 	\label{figbands}
 \end{figure}

The resulting set of zeros computed with $\Lambda = .999$, for a few representative values of $L$ are shown in Figure (\ref{figbands}). They display continuous and non intersecting lines - the {\it Floquet spectra}  $\left \{ \eta=\chi_b(\omega)\right \}_{b=1}^{B(L)}$   where $B(L)$ denotes  their number. Some lines appear to end abruptly and they are replaces by isolates points or intervals. This is the result of insufficient numerical resolution as explained previously. This is demonstrated  for $L=1.75$, where the lines  do not continue beyond  $\omega=0.8 \pi$ (see Figure (\ref{figbands}) first frame of the middle line.) However,  they  are recovered by the high resolution calculations shown in  Figure (\ref{highres}).

Based on numerous computations, a coherent systematics of the networks of  Floquet spectra, and their dependence on the relevant parameters can be obtained:  For any given ${\sqrt \frac{2}{3}}< \Lambda<1$,
the Floquet spectral curves $\eta = \chi_b (\omega)$ are monotonic decreasing functions of $\omega$. They start and end on the boundaries of their rectangular domains of definition, however, not necessarily on parallel edges. For the Floquet spectrum $\eta=\chi_b (\omega)$, the starting position will be denoted by $(\omega_s,\eta_s)_b$ and the ending position by  $(\omega_e,\eta_e)_b$, with $\omega_s \le \omega _e$ and $\eta_s\ge \eta_e$.

The networks  of Floquet spectra change dramatically with $L$. This development is demonstrated in Figure (\ref{figbands}):

For the lower values of $L$, the spectral curves are restricted to the domain $\omega<\frac{\pi}{2}$. Some curves start at  $(\omega_s =0,\eta_s <\frac{1}{2})$ and  terminate at $(\omega_e < \frac{\pi}{2},\eta_e=0)$.  Most of the curves are almost perpendicular to the $\omega$ axis, with $(\omega_s >0,\eta_s=\frac{1}{2})$ and $(\omega_e,\eta_e=0)$ where $\omega_e>\omega_s$ and  $(\omega_e-\omega_s) \ll \pi$. For all curves, $\omega_e < \frac{\pi}{2}$.

When $L$ increases to moderate values ($L  \gtrsim 1$), the mean slopes of the curves is decreasing, some curves start at  $(\omega_s=0,\eta_s<\frac{1}{2})$ and end either at $(\omega_e< \pi,\eta_e=0)$ or at $(\omega_e= \pi,\eta_e>0)$. Some other lines connect the points $(\omega_s>0, \eta_s=\frac{1}{2})$ with $(\omega_e=\pi, \eta_e >0)$.  As $L$ increases the number of the later kind decreases. At the same time, the number of curves $B(L)$ decreases too. In the present case, it starts with $B(L=.5)=12$, and as the flattening of the curves continues it decreases, with  $B(L>1.75)=6 $. For  $L>5$, almost all the lines are quite parallel to the $\omega$ axis.

For asymptotically large $L$ the  alignment of the Floquet spectra become almost parallel to the $\omega$ axis and the spectral bands $(\eta_s-\eta_e)$ become narrower, converging to the spectra of the corresponding single $\delta$-potential case. This happens for very large $L$ ($\approx  50$) for the case demonstrated numerically). The lowest curve which connects the $\omega$ and $\eta$ axes (see lowest frames in Figure (\ref{figbands})) persists to exist until  the very large $L$ domains is reached.

The features of the Floquet spectra discussed above are typical for other values of $\Lambda$. The most interesting feature is the dependence of the Floquet spectra  on the parameter $L$: A transition from a group of approximately vertical and parallel lines at low $L$  values, to a series of horizontal  parallel lines for large $L$. This is accompanied by a monotonic decrease of the number of lines $B(L)$. The transition is continuous and occurs in an $L$ interval which depends on the interaction strength $\Lambda$. When $\Lambda$ gets closer to $1$ the spectra become denser and the features which characterize the extreme  low and high $L$ values become more acute. The transition has severe effect on the spectral bands of the periodic Hamiltonian. The spectrum consists of the union of the intervals $\bigcup_{b=1}^{B(L)}[\eta_e,\eta_s]_b$ (the bands). For low $L$ the spectrum is continuous on the entire $\eta$ domain and it is  $B(L)$ times degenerate. For large $L$ the spectral bands narrow down and converge to points in the limit. Such a transition has severe implications to the transmission properties (conductance) of this model.

This spectral transition phenomenon can be discussed and explained further within the  WKB  framework which was developed in subsection \ref {subsecwkb}.

Starting with the discrete recursion relation  (\ref {kprecursion}), one can  approximate it  by substituting
\begin{equation}
D_m= q_m \prod_{r=0}^m [\Lambda u_r \sqrt {\frac{r+\frac{1}{2}}{r+\eta}}]
\end{equation}
and neglecting terms of order $\frac{1}{m^2}$. This gives,
\begin{equation}
q_{m+1}=\left[\frac{1-\frac{\cos\omega}{\cosh \sqrt{m+1+\eta}L}}
 {\tanh\sqrt{m+1+\eta}L}\right]\frac{2}{\Lambda} \sqrt {\frac{m+1+\eta}{m+1+\frac{1}{2}}} q_m -q_{m-1} \ .
 \label{kpdiscrete}
\end{equation}
where the factor in the square brackets above contains all the changes which are brought about by going from a single $\delta$ to a periodic $\delta$ interaction. One can justify approximating this discrete equation by an ODE relying on the same arguments as used previously to derive (\ref{wkb1}).  They are based on the fact  that
the effective potential
\begin{equation}
V(m) =\left[\frac{1-\frac{\cos\omega}{\cosh \sqrt{m+1+\eta}L}}
 {\tanh\sqrt{m+1+\eta}L}\right]\frac{2}{\Lambda} \sqrt {\frac{m+1+\eta}{m+1+\frac{1}{2}}}\ .
 \label {kppotential}
\end{equation}
is smooth in the interior of the rectangular domain of interest in the $(\omega,\eta)$ plane.

 The continuous version of (\ref{kpdiscrete})is,
 \begin{equation}
 \hspace{-20mm}
-\frac{{\partial}^2 q(m)}{{\partial} m^2} +\left[\left[\frac{1-\frac{\cos\omega}{\cosh \sqrt{m+1+\eta}L}}
 {\tanh\sqrt{m+1+\eta}L}\right]\frac{2}{\Lambda} \sqrt {\frac{m+1+\eta}{m+1+\frac{1}{2}}}\ - 2\  \right ]\ q(m) = 0\ ,  \ \  m \geq 0 .
\label {kpode}
\end{equation}
where $q(m)$ stands for a function of $m\in \mathbb{R^+}$ and assumes the values $q_m$ for integer $m$. This equation can be thought of as a Schr\"odinger equation describing a particle of "energy" $2$ on the positive half line, subject to the potential $V(m)$.
The boundary condition at $m=0$ is derived from the the fact that $D_0=2(1-\frac{\cos \omega}{\cosh (\sqrt {\eta}L)})$ :
\begin{equation}
q(0)=\frac{ \sqrt{ 2\eta}}{\tanh ( \sqrt{\eta}L)}  \frac{2}{\Lambda}(1-\frac{\cos \omega}{\cosh ({\sqrt \eta}L)})\ .
\label{kpbcond}
\end{equation}
For the sake of notational simplicity, the parametric dependence of both $V(m)$ and $q(m)$ on $\eta, \omega, L, \Lambda$ will be omitted.

A few properties of $V(m)$ and of the boundary condition (\ref {kpbcond}) will now be derived. The potential plays a crucial role in the discussion of the Floquet spectrum.  The WKB eigenfunction of interest here is the one with energy $2$. The number of its zeroes, according to Sturm's oscillation theorem is given by the number of eigenfunctions with lower energies. Following the discussion in the previous section,  this number relates directly to the number of $\eta$ values at which the secular function vanishes. We see therefore that the WKB method enables the deduction of the properties of the spectral bands from the functional shape of the scaled potential. Of prime importance in the semiclassical analysis is the action integral
\begin{equation}
\sigma(L,\omega,\eta) = \int_{m_0}^{m_t}  \sqrt {2-V(\mu)} {\rm d}\mu\ ,
\end{equation}
where $m_0$ and $m_t, (m_t>m_0)$ are the classical turning points with $m_0=0$ if the potential intersects the $0$ lines only once. The action integral is proportional to the number of oscillations of the wave function in the interval $(m_0,m_t)$, which can be translated to the number of $\eta$ spectral values which appear for a given Floquet phase $\omega$.

The boundary condition (\ref {kpbcond}) is regular for the entire range of $\eta$ and $\omega$. It becomes independent of $\omega$ in the limit of large $L$ values.

In the sequel, it is convenient to write $V(m)$ in a different form:
\begin{equation}
\hspace{-15mm}
V(m)=\left [ \tanh (\frac{L}{2}\sqrt{m+1+\eta}) +\frac{2\sin^2\frac{\omega}{2}}{\sinh (L\sqrt{m+1+\eta})}\right ] \left( \frac{2}{\Lambda} \sqrt {\frac{m+1+\eta}{m+1+\frac{1}{2}}}\right )\ .
\label {kppotential1}
\end{equation}

The following properties of the potential $V(m)$ may be used to explain the main features of the numerically computed Floquet spectra :

\noindent ${\it i}.$ For large $m$ (and any $L>0$) the square bracket takes the value $1+\mathcal{O}(e^{-\sqrt{m} L})$. The circular brackets consists of  the potential of a single $\delta$ function. For large $m$ it approached $\frac{2}{\Lambda}$ which for $\Lambda <1$ exceeds $2$. Therefore, for both the single and periodic  $\delta$, the WKB wave functions exhibit the same asymptotic behavior for large $m$ values.

\noindent ${\it ii}.$ For large $L$ the potentials  (\ref {kppotential1}) approaches its single $\delta$ counterpart for all $m$ and the boundary conditions  at $m=0$ also coincide. Hence the Floquet spectrum for large $L$ is expected to consists of curves which converge to parallel lines at $\eta$ values which coincide with the  spectrum of the corresponding single $\delta$ potential. This explains the behavior of the Floquet spectra in the asymptotically large $L$ domain,  as illustrated for $L=5$ in the righmost frame in the lowest line of Figure (\ref{figbands}. The value $L=5$ is not large enough, and hence the deviation of the lowest Floquet spectrum. This phenomenon will be discussed in point ${\it v.}$ below.

\noindent ${\it iii}.$  At $\omega =0$  the single $\delta$ potential is modified for $m$ in the range $0\le m < \frac{1}{L^2} $. This is effective in the domain  $L< 1$ where indeed the Floquet spectrum at $\omega =0$ is very different from the single $\delta$ potential. (see  Figure (\ref{figbands}) top left frame for $L=.5$). The changes of the boundary conditions at $m=0$ also contribute to the difference, but the similarity is restored as $L$ increases.

\noindent ${\it iv}.$ Consider the parameter domain $L<<1 $ and $\omega > 0$. Expanding the potential (\ref{kppotential1}) formally with respect to
$L\sqrt{m+1+\eta}$ one gets,
\begin{equation}
\hspace{-25mm}
V(m)= \frac{2}{L\Lambda}\frac{1}{\sqrt{m+\frac{3}{2}}}\left[ 2\sin^2\frac{\omega}{2} + (\frac{1}{2}-\frac{\sin^2\frac{\omega}{2}}{3})(L\sqrt{m+1+\eta})^2 + \mathcal{O}((L\sqrt{m+1+\eta})^4) \right]
\end{equation}
The leading term in the approximate potential is independent of $\eta$ and its dependence on $m$ is simple. The classical turning point for this potential is $m_t(\omega=0)=\frac{2 \sin^2\frac{\omega}{2}}{\Lambda L}-\frac{3}{2}. $ Hence, to support an eigenfunction $m_t(\omega=0)$ must exceed $0$ so that $\sin^2\frac{\omega}{2}> \frac{3}{4}\Lambda L$. This is a bound from bellow on the $\omega$ range for which this approximation is allowed. The range of $m$ is bounded from above by the requirement that the expansion parameter $L\sqrt{m+1+\eta}$ is small. Substituting the expression for $m_t(\omega=0)$ derived above shows that  this value (and hence for all $m<m_t(\omega=0)$) is of order $\sin\frac{\omega}{2}\sqrt{2L}\ <\ 1$. Hence, the leading term in the potential suffices.  This result has very important consequences for the Floquet spectrum in the domain of $\omega$ where it is valid:  the weak dependence of the potential on $\eta$, implies that the Floquet spectrua  for small $L$ are almost parallel to the $\eta$ axis in the domain $3\sqrt{\Lambda L}<\omega < \frac{\pi}{2} $. This provides a clear explanation for the occurrence of perpendicular Floquet spectra  at the low $L$ domain, which appear in a domain which start away from $\omega=0$.  (see  Figure (\ref{figbands}) top left frame for $L=.5$.

\noindent {{\it v.} For $L>1$ and $\omega < \frac{\pi}{2}$ one can approximate the hyperbolic functions in the potential to leading order in $e^{-\sqrt{m+1+\eta} L}$
\begin{equation}
V(m) \approx \left[ 1-2e^{-\sqrt{m+1+\eta} L}(1-2\sin ^2\frac{w}{2})\right ] \frac{2}{\Lambda}\sqrt{\frac{m+1+\eta}{m+1+\frac{1}{2}}}\ .
\end{equation}
In order that this potential would bind a wave function with "energy" $2$, its value at $m=0$ should be smaller than $2$. This happens for $\eta <\eta_{max}$ which was defined in the preceding section (\ref {etamax}). Since for large $m$, $V(m)$ reaches $\frac{2}{\Lambda}$ which is larger than $2$, there exists  a classical turning point. Its difference from  the turning point $m_t$ of the single $\delta$ interaction, is bounded in the interval
\begin{equation}
\hspace{-15mm}
0\le \delta_m \le \frac{2}{3} (\frac{\frac{\Lambda^2}{2}-\eta}{1-\Lambda^2})^2 e^{-L\sqrt{1+m_t}} (1-2\sin ^2\frac{w}{2})\  \ \ ({\rm with}\
m_t=\frac{\frac{\Lambda^2}{2}-\eta}{1-\Lambda^2}-1 ) .
\end{equation}
Clearly, the only way to increase the range of $m$ where $V(m) <2$ is by decreasing $\eta$ towards $0$ so that the large factor $\frac{1}{(1-\Lambda^2)^2}$ would counteract the exponential factor $e^{-L\sqrt{1+m_t}}$ in the expression for $\delta_m$. This could also be achieved by decreasing $\omega$. At the same time one should observe that the exponential factor is less effective due to the fact that $L$ is multiplied by $\sqrt{m_t}$. This mechanism may cause a Floquet spectrum to appear in the low $\eta$ and low $\omega$ domain even for moderate $L$.

\begin{figure}
  \includegraphics[width=.33\textwidth]{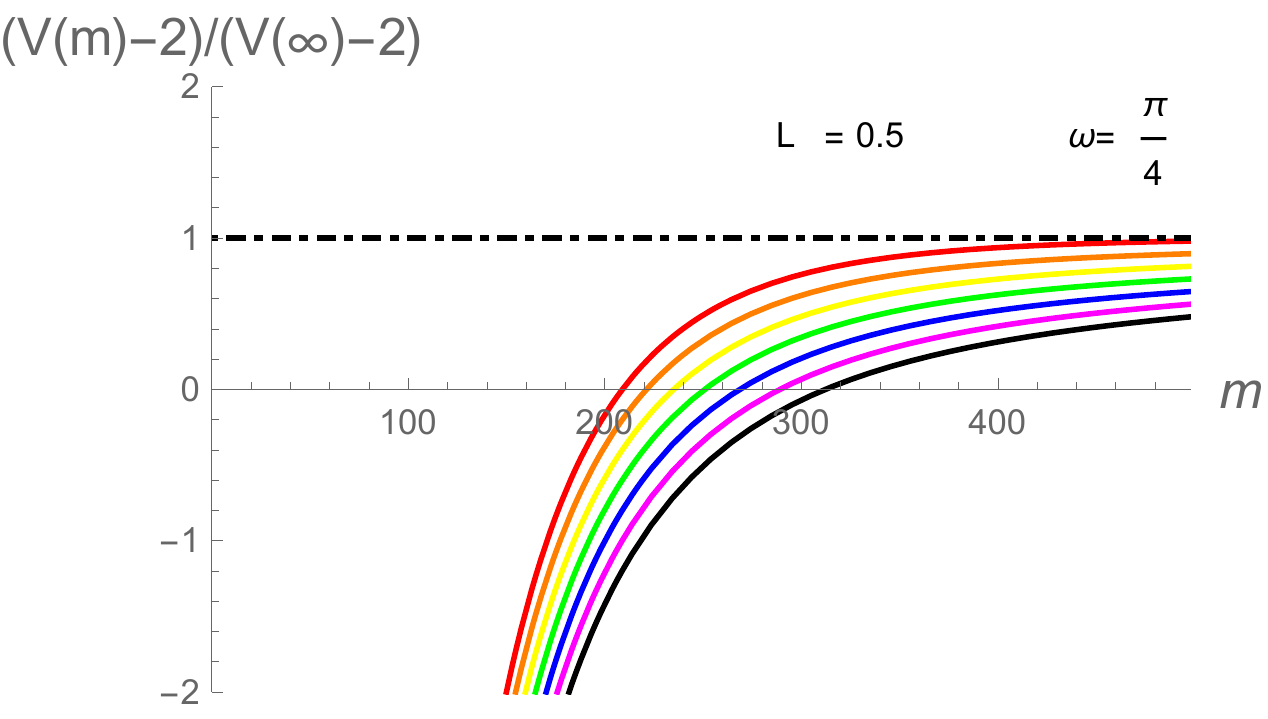}\hfill
  \includegraphics[width=.33\textwidth]{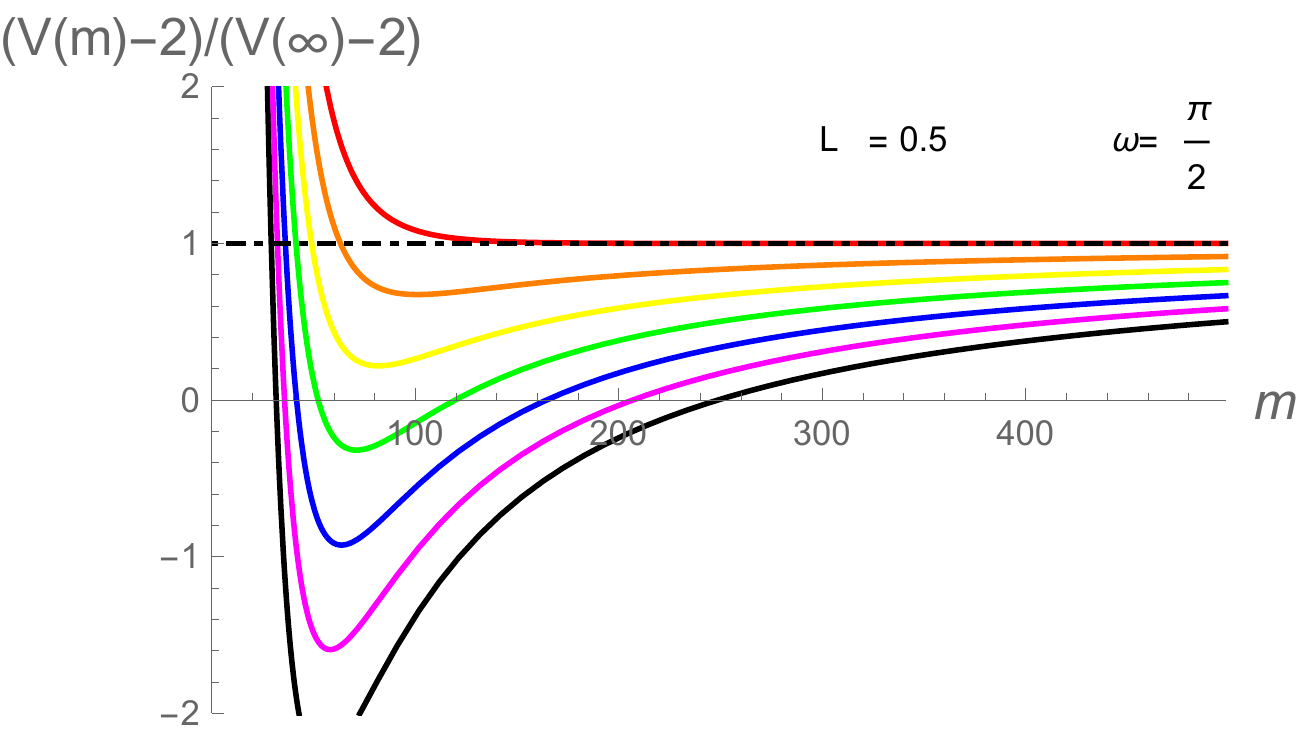}\hfill
  \includegraphics[width=.33\textwidth]{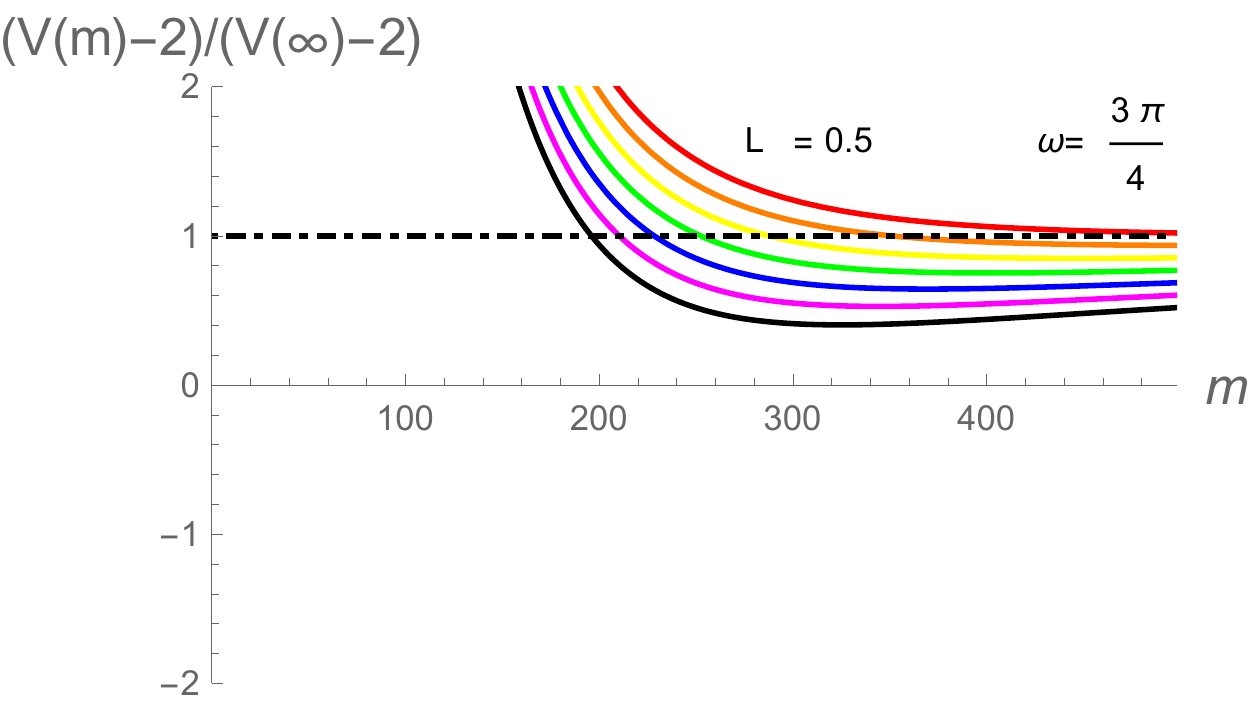}\hfill
 \includegraphics[width=.33\textwidth]{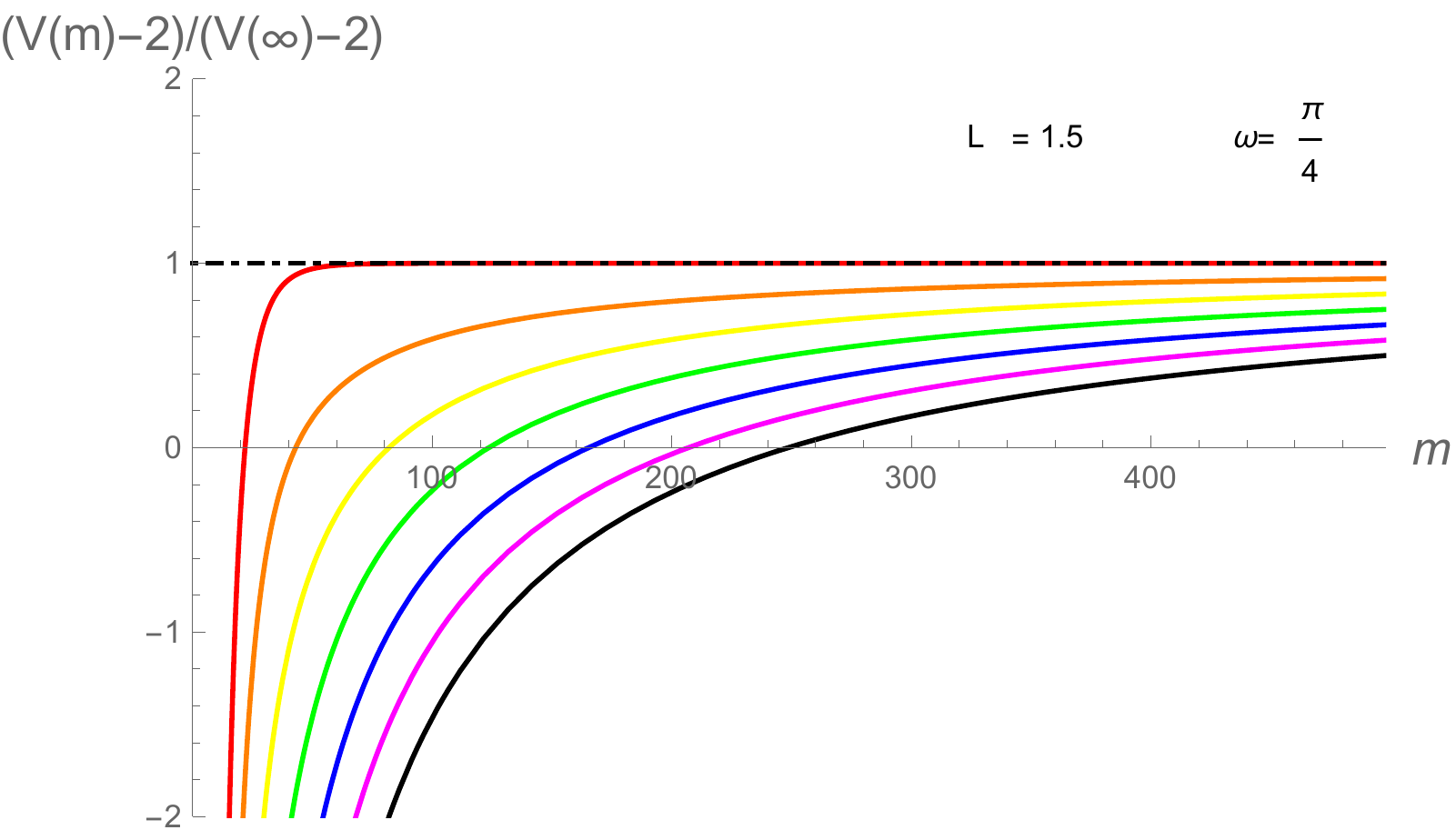}\hfill
 \includegraphics[width=.33\textwidth]{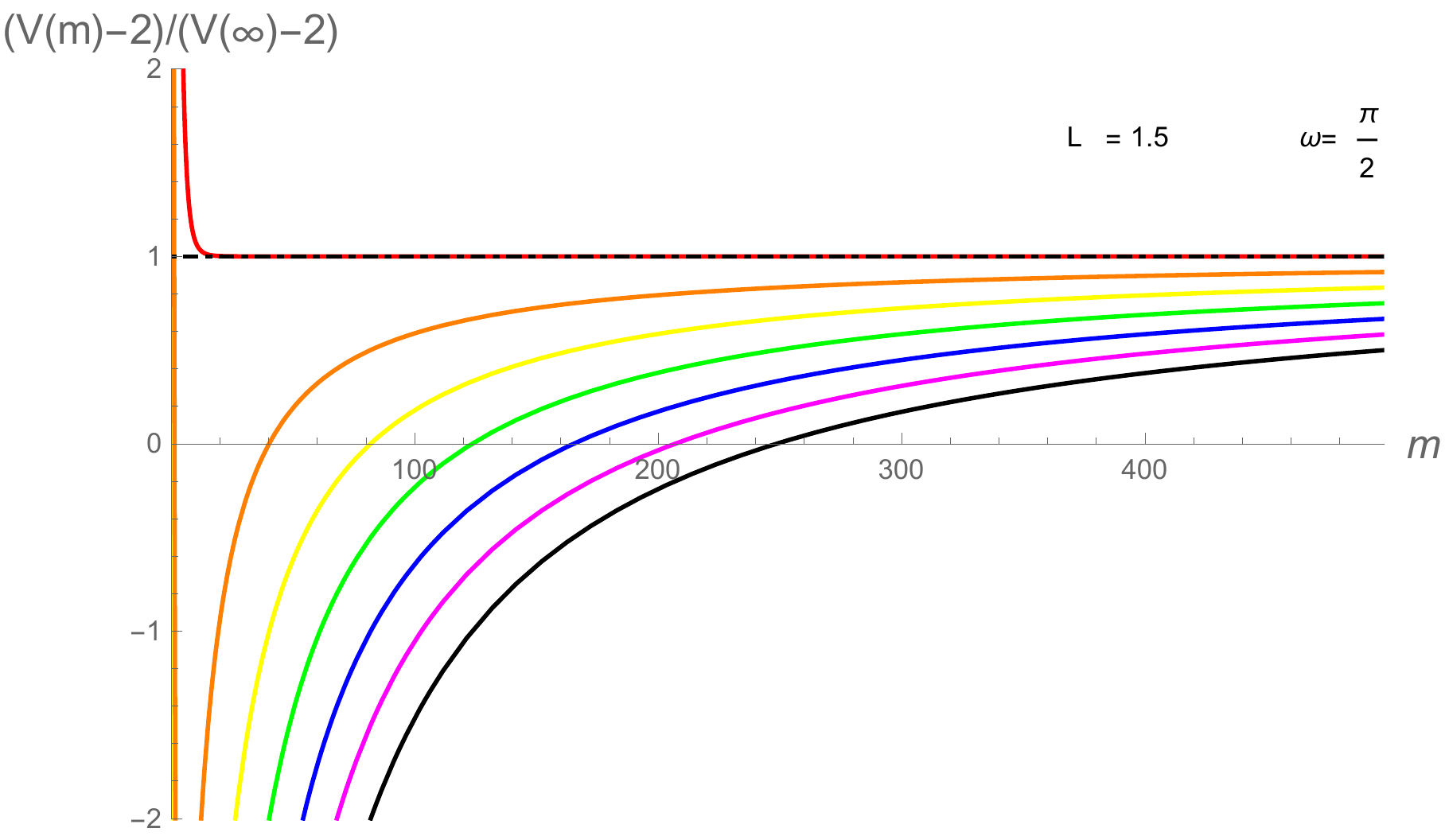}\hfill
 \includegraphics[width=.33\textwidth]{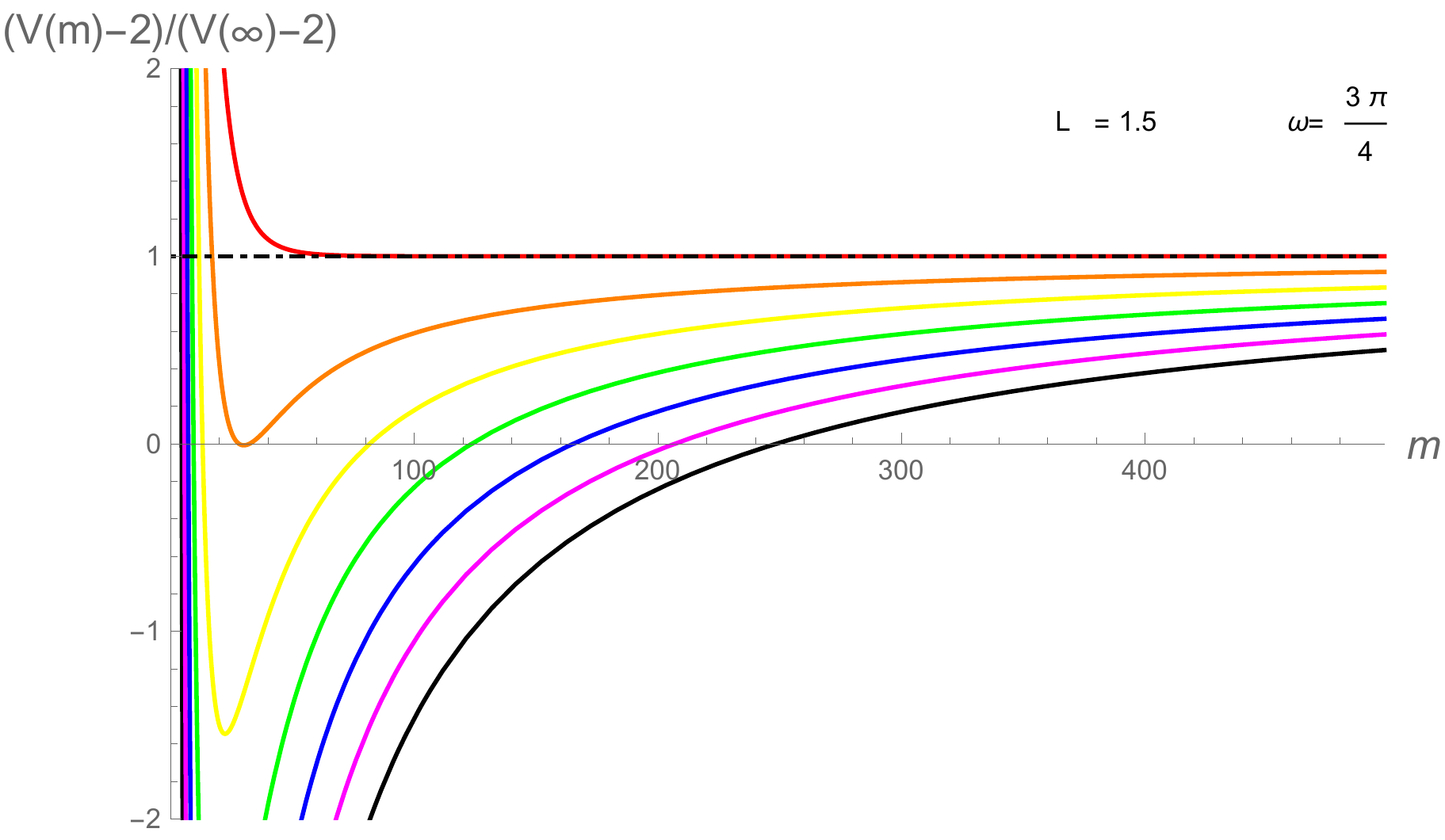}\hfill
\includegraphics[width=.33\textwidth]{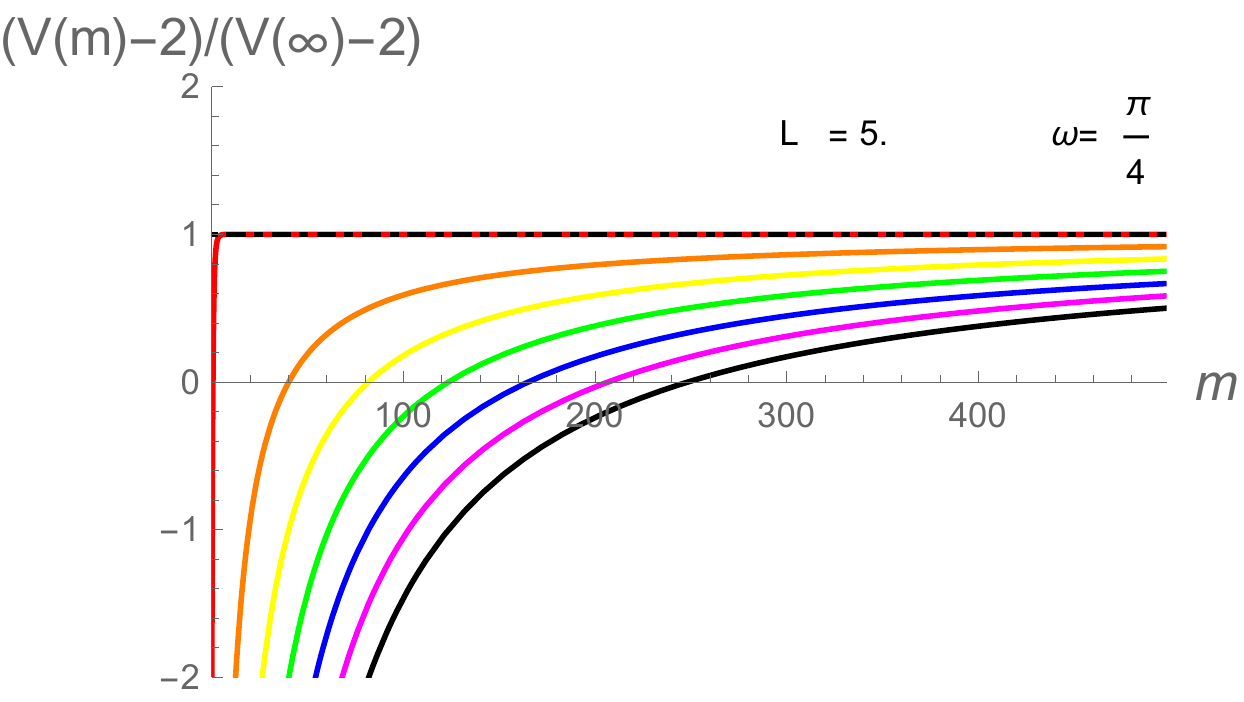}\hfill
\includegraphics[width=.33\textwidth]{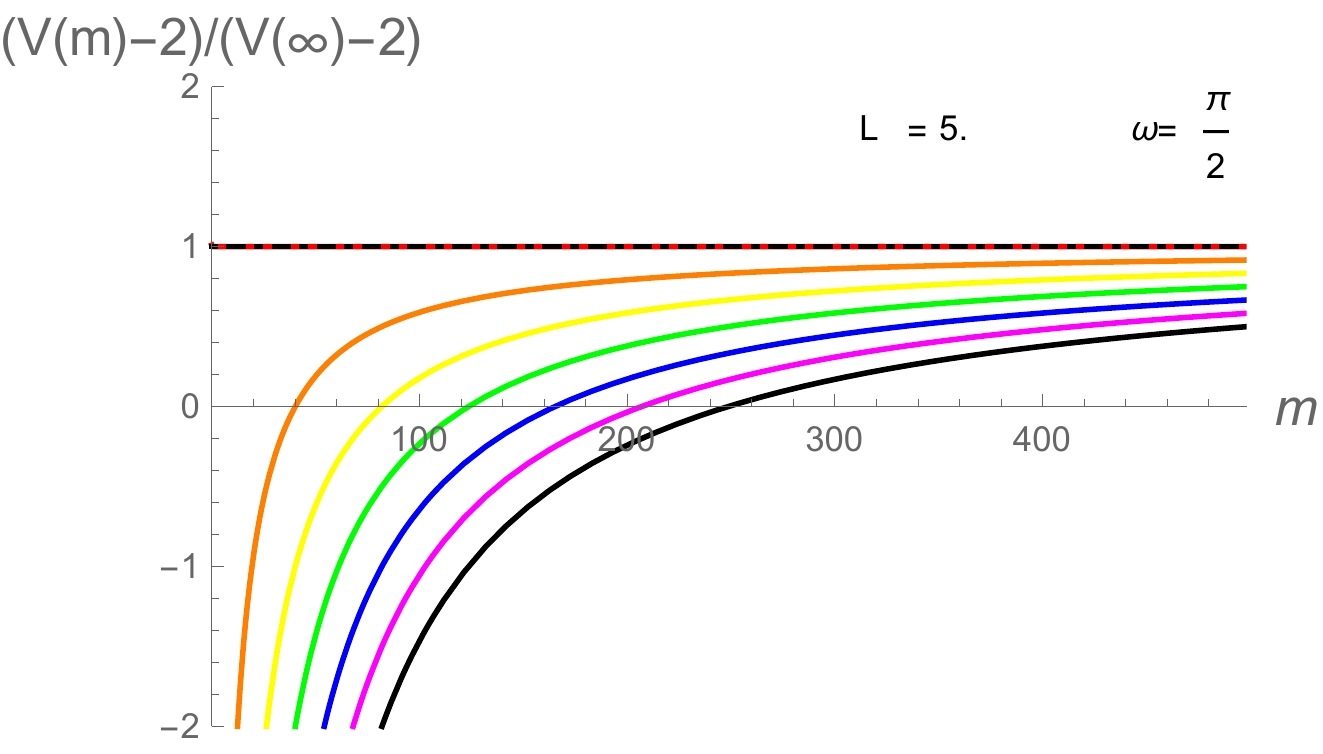}\hfill
\includegraphics[width=.33\textwidth]{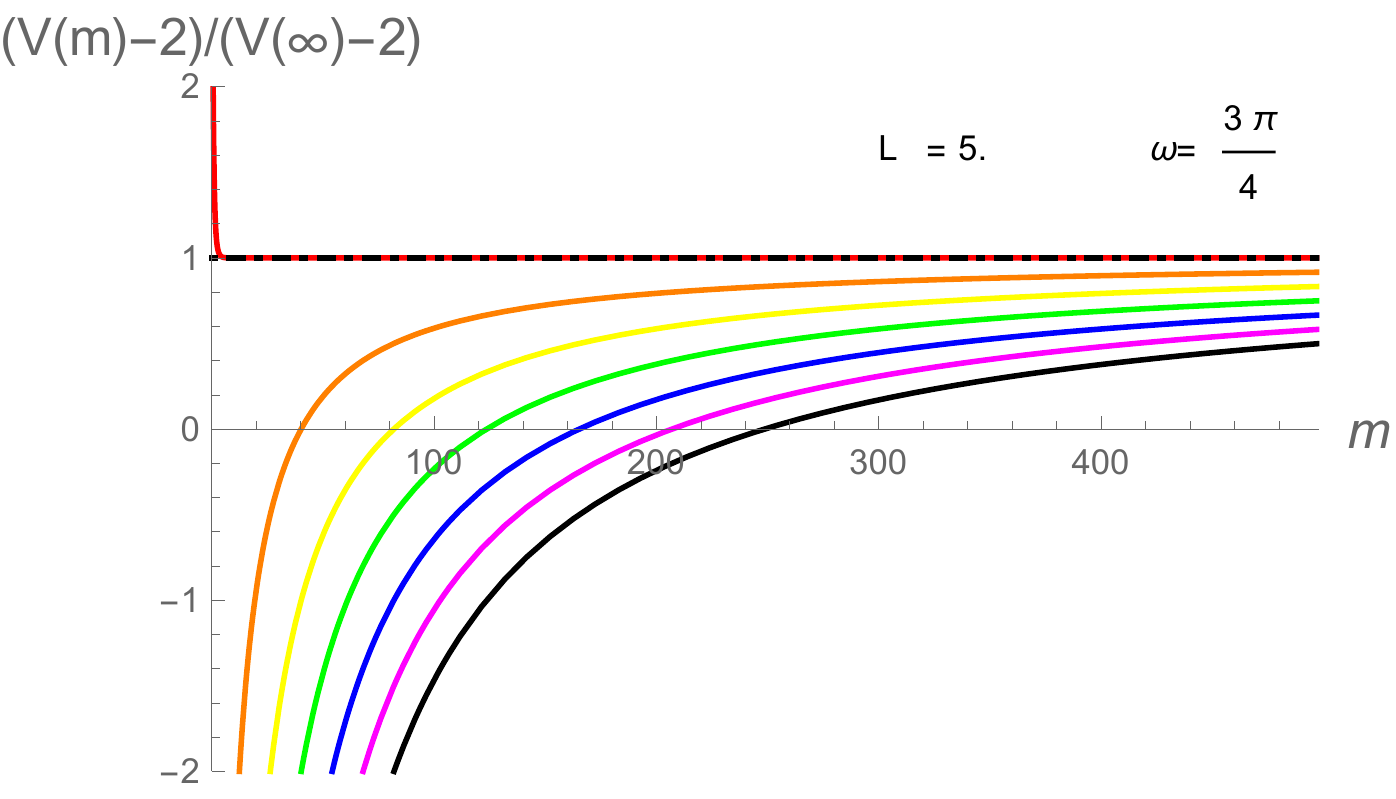}\hfill
 	\caption{The scaled potential $\frac{V(m)-2}{V(\infty)-2}$ as a function of $m$ for three $L$ values : $L=0.5,1.5,5.0 $ and for $\omega =\frac{\pi}{4}, \frac{\pi}{2},\frac{3 \pi}{4}$.  In each
 frame the different colored curves correspond  to seven $\eta$  values : $\eta=0.$ (black), $\eta=1/12$ (violet), $\eta=2/12$ (blue), $\eta=3/12$ (green), $\eta=4/12$ (yellow), $\eta=5/12$ (orange), $\eta=1/2$ (red) The horizontal  dot-dashed line at $1$ is the common asymptotic value of the normalized potentials for all the cases.}
 \label{kpottt}
 \end{figure}

Quite a detailed survey of the dependence of $V(m)$ on the various parameters is provided  in  Figure (\ref{kpottt}).  To facilitate the comparison between the various cases, the quantity plotted is the "scaled potential" $\frac{V(m)-2}{V(m=\infty)-2}$ where $V(m=\infty ) =\frac{2}{\Lambda}$  is the asymptotic  value of the potential and is independent of the parameters $\eta, \omega$ and $L$. The "energy" 2 is subtracted because this difference dictates the rate of growth of the asymptotic wave function. In this representation, the classical turning points are the zeros of the scaled potential, and the domain of $m$ values where the scaled potential is negative is the potential well where eigenfunctions with energy 2 are trapped. Outside, the eigenfunctions decay or increase exponentially. The figures are truncated at the scaled potential $= -2$.   For all potential parameters, except for $\eta=0$  the values of  $V(m=0)$ are finite. However,  $V(m=0,\eta\rightarrow 0)$ diverges as $\eta^{-\frac{1}{2}}$. Therefore only a finite number of eigenfunctions can be trapped. The eigenfunction of interest here is the one with energy $2$. Following the discussion in the previous section, the number of eigenstates with energy less than $2$ counts the number of zeros, which, in turn relates directly to the number of $\eta$ values at which the secular function vanishes.

Figure (\ref{kpottt})is arranged in the following way. Each row is computed for a single value of $L$: $L=0.5$ represents the lower range of $L$ where the spectral bands are almost perpendicular to the $\omega$ axis. The transition range is represented by $L=1.5$ and the data occupies the second row. The range of large  $L$ values  where the spectral bands are parallel to the $\omega$ axis is represented by $L=5.$ at the lowest row. The frames at each of the rows are computed (from left to right) for $\omega = \frac{\pi}{4}, \frac{\pi}{2},\frac{3 \pi}{4}$. In each frame, the scaled potential is computed for 7 values of $\eta$ ranging at equal intervals from $0$ to $0.5$, and are colored according to the spectrum of visible light, with  black for $\eta=0$.

All the potential lines  for $ L=\ 0.5$ are consistent with the observations made before. For $\omega = \frac{\pi}{4} \ \ \ $ the potentials  are monotonically increasing from negative values to $1$. They are close, getting closer as $m$ decreases, and all the potentials lines intersect the $0$ axis. (For smaller $\omega$ the lines are getting nearer). The action integrals are large and  depend only mildly on  $\eta$ since the potentials become closer as $m$ decrease, and the bulk of the integrals is due to the low $m$ domain. This is consistent with the observed behavior of the spectrum at the interval $\omega <\frac{\pi}{2}$, and with the quantitative discussion provided in point $ {\it iii.} $   above .
For $ \omega = \frac{\pi}{2} \ \ \ $  the potential has a minimum and from there on it grows steeply. The only potential lines which intersect the $0$ axis are with $\eta < 1/3$ (the black, violet, blue and green lines). It is expected therefore that in the middle range of $\omega$ the spectral band will avoid the low $\eta$ domain. The action integral will be dramatically decreased, resulting in only a few spectral bands in this $\omega $ range.
For $ \omega = \frac{3\pi}{4} \ \ \ $ none of the potential lines intersects the $0$ line. Therefore this $\omega$ range will not support any spectral band. The above qualitative discussion is consistent with the data presented in the upper leftmost frame of Figure (\ref{figbands}).

In the intermediate and large $L$ domains the variation of the potentials corresponding to different $\eta$ and $\omega$ values are not as large as in the small $L$ domain. They correspond to the slow variation of the Floquet spectra .

Clearly,the Floquet spectral transition is due to the corresponding transition in the potential $V(m)$.
The arguments above provide a qualitative description of the complex behavior observed in the numerical solution of the original recursive computation of the secular equation. The most important feature - the transition of the energy bands $[\eta_e,\eta_s]_b$  from covering the entire energy domain for low $L$ to narrow domains converging to points as $L\rightarrow \infty$  is explained.   A detailed WKB analysis is complicated since the action integrals cannot be performed analytically, and the uniform approximation in terms of the Airy functions is not valid when two turning points exist. This is also the reason why the  variation of the number of Floquet spectra, $B(L)$ cannot be obtained.  A more precise  treatment is certainly called for.

\section{Summary and Conclusions}
The simple looking model of a particle in a quadratic channel with $\delta$ interaction  was introduced some 20 years ago \cite {irrev}, and since then it continues to offer a surprising store of new and interesting problems. This is true also for the present attempt to replace the single $\delta$ by a periodic array of $\delta$ interactions thus turning the system into a Kronig-Penney like model. In the present series we have shown that the spectral transition which occurs for the single delta model persists also for the periodic case, however, new phenomena emerge when one studies the underlying dynamics as shown in (I). In the present paper the surprises arise in the discussion of the spectral bands in the sub-critical domain, and in particular on the strong dependence of the bands on the periodicity interval $L$. The numerical results and the  explanations based on the WKB analysis, reveal the intricacy of the system but leave a lot to be desired, and in particular requires more research.

\section*{Acknowledgements}
This work was concluded in the University of Bath, where the author was  nominated a David Parkin visiting professor in the department of mathematical studies. Thanks for the kind hospitality are very much due. This work started as a collaboration with Professor Italo Guarneri, and his  ideas and critical comments  where essential throughout. Thanks Italo! Thanks are also due to Professor Sven Gnutzmann who accompanied this project with a study of multi-mode quantum graphs.His interest and suggestions were invaluable.

\end{document}